\documentclass[twocolumn]{aastex6}

\pdfoutput=1

\newcommand{\mearth}{$M_{\oplus}$}

\newcommand{\MOST}{\textit{MOST}}
\newcommand{\HATS}{HATSouth}
\newcommand{\Kepler}{\textit{Kepler}}
\newcommand{\multi}{{\sc MultiNest}}
\newcommand{\FBEYE}{{\tt FBEYE}}
\newcommand{\forecaster}{{\tt Forecaster}}
\newcommand{\ms}{m\,s$^{-1}$}
\newcommand{\urllink}{https://github.com/CoolWorlds/Proxima}

\usepackage{amsmath}
\usepackage{amsfonts}
\usepackage{amssymb}
\usepackage{wasysym}
\usepackage{comment}

\begin{document}


\title{No Conclusive Evidence for Transits of Proxima b in \MOST\ photometry}


\author{
David M. Kipping\altaffilmark{1}, Chris Cameron\altaffilmark{2}, Joel D. Hartman\altaffilmark{3},
James R. A. Davenport\altaffilmark{4,5}, Jaymie M. Matthews\altaffilmark{6},\\ Dimitar Sasselov\altaffilmark{7},
Jason Rowe\altaffilmark{8}, Robert J. Siverd\altaffilmark{9}, Jingjing Chen\altaffilmark{1}, Emily Sandford\altaffilmark{1},
G\'asp\'ar \'A. Bakos\altaffilmark{3,10,11},\\Andr\'es Jord\'an\altaffilmark{12,13}, Daniel Bayliss\altaffilmark{14},
Thomas Henning\altaffilmark{15}, Luigi Mancini\altaffilmark{15}, Kaloyan Penev\altaffilmark{3},
Zoltan Csubry\altaffilmark{3},\\ Waqas Bhatti\altaffilmark{3}, Joao Da Silva Bento\altaffilmark{16},
David B. Guenther\altaffilmark{17}, Rainer Kuschnig\altaffilmark{18}, Anthony F. J. Moffat\altaffilmark{8},\\
Slavek M. Rucinski\altaffilmark{19}, Werner W. Weiss\altaffilmark{20}
}

\affil{dkipping@astro.columbia.edu}

\altaffiltext{1}{Department of Astronomy, Columbia University, 550 W 120th St., New York, NY 10027}
\altaffiltext{2}{Department of Mathematics, Physics \& Geology, Cape Breton University, 1250 Grand Lake Road, Sydney, Nova Scotia, B1P 6L2, Canada}
\altaffiltext{3}{Department of Astrophysical Sciences, 4 Ivy Ln., Princeton, NJ 08544}
\altaffiltext{4}{Department of Physics \& Astronomy, Western Washington University, 516 High St., Bellingham, WA 98225}
\altaffiltext{5}{NSF Astronomy \& Astrophysics Postdoctoral Fellow}
\altaffiltext{6}{Department of Physics \& Astronomy, University of British Columbia, 6224 Agricultural Road, Vancouver, BC, V6T 1Z1, Canada}
\altaffiltext{7}{Harvard-Smithsonian Center for Astrophysics, 60 Garden Street, Cambridge, MA 02138}
\altaffiltext{8}{Observatoire Astronomque du Mont M\'egantic, D\'epartment de Physique, Universit\'e de Montr\'eal C. P. 6128, Succursale, Centre-Ville, Montr\'eal, QC, H3C 3J7, Canada}
\altaffiltext{9}{Las Cumbres Observatory Global Telescope Network, 6740 Cortona Drive, Suite 102, Santa Barbara, CA 93117}
\altaffiltext{10}{Packard Fellow}
\altaffiltext{11}{Alfred P. Sloan Fellow}
\altaffiltext{12}{Instituto de Astrof\'isica, Universidad Cat\'olica de Chile, Av. Vicu\~na Mackenna 4860, 7820436 Macul, Santiago, Chile}
\altaffiltext{13}{Millenium Institute of Astrophysics, Av. Vicu\~na Mackenna 4860, 7820436, Santiago, Chile}
\altaffiltext{14}{Observatoire Astronomique de Universite de Gen\'eve, 51 ch. des Maillettes, 1290 Versoix, Switzerland}
\altaffiltext{15}{Max Plank Institute for Astronomy, K\"onigstuhl 17, 69117 Heidelberg, Germany}
\altaffiltext{16}{Research School of Astronomy \& Astrophysics, Mnt. Stromlo Obs., Australian National Univ., Cotter Road, Weston, ACT 2611, Australia}
\altaffiltext{17}{Institute for Computational Astrophysics, Department of Astronomy and Physics, Saint Mary's University, Halifax, NS B3H 3C3, Canada}
\altaffiltext{18}{Institut für Astronomie, Universit\"at Wien T\"urkenschanzstrasse 17, A1180 Wien, Austria}
\altaffiltext{19}{Department of Astronomy and Astrophysics, University of Toronto, Toronto, ON M5S 3H4, Canada}
\altaffiltext{20}{Institute for Astrophysics, University of Vienna, Universit\"atsring 1, 1010 Vienna, Austria}

\begin{abstract}

The analysis of Proxima Centauri's radial velocities recently led \citet{nature}
to claim the presence of a low mass planet orbiting the Sun's nearest star once
every 11.2\,days. Although the a-priori probability that Proxima b transits its
parent star is just 1.5\%, the potential impact of such a discovery would be
considerable. Independent of recent radial velocity efforts, we observed
Proxima Centauri for 12.5\,days in 2014 and 31\,days in 2015 with the
\MOST\ space telescope. We report here that we cannot make a compelling case that
Proxima b transits in our precise photometric time series. Imposing an 
informative prior on the period and phase, we do detect a candidate signal with
the expected depth. However, perturbing the phase prior across 100 evenly 
spaced intervals reveals one strong false-positive and one weaker instance. We
estimate a false-positive rate of at least a few percent and a much higher 
false-negative rate of $20$-$40$\%, likely caused by the very high flare rate 
of Proxima Centauri. Comparing our candidate signal to \HATS\ ground-based
photometry reveals that the signal is somewhat, but not conclusively, disfavored
(1-2\,$\sigma$) leading us to argue that the signal is most likely spurious. We
expect that infrared photometric follow-up could more conclusively test the
existence of this candidate signal, owing to the suppression of flare activity and
the impressive infrared brightness of the parent star.

\end{abstract}

\keywords{planetary systems --- stars: individual (Proxima Centauri) --- techniques: photometric\\}



\section{INTRODUCTION}
\label{sec:intro}

Proxima Centauri is the Sun's nearest stellar neighbor at a distance of 1.295
parsecs \citep{2007A+A...474..653V}. Despite this, Proxima's late spectral type
(M5.5; \citealt{1991AJ....101..662B}) makes it too faint to be seen by the 
naked eye ($V=11.1$; \citealt{2014AJ....147...21J}), elucidating why this is 
not the easiest target in the search for extrasolar planets. This
challenge is exacerbated by the activity of Proxima itself, being a classic flare
star \citep{2004ApJ...612.1140C}.

Nevertheless, Proxima is one of the best studied low-mass stars and its
diminutive mass offers an enhanced radial velocity semi-amplitude, $K$, scaling
as $M_{\star}^{-2/3}$. Accordingly, translating the same planet from the Sun to
Proxima would cause $K$ to increase by a factor of four. Early radial velocity
campaigns, such as those of \citet{2008A+A...488.1149E} and
\citet{2009A+A...505..859Z} found no signals at the few \ms\ level, ruling out
Super-Earths in the habitable-zone.

The hunt for planets around our nearest star fell to the sidelines in the
following years, notably during the era of NASA's \textit{Kepler Mission}.
With thousands of planetary candidates detections pouring in
\citep{2013ApJS..204...24B}, the exoplanet community reasonably focussed
on these immediate discoveries. Although only a few thousand M-dwarfs were
observed by \Kepler\ (out of $\sim$200,000 targets), the \Kepler\ results
ultimately rekindled our team's interest in the prospect of planets around 
Proxima.

First, the discovery of a planetary system around one of \Kepler's lowest mass
stars, Kepler-42 (M5 dwarf), \citet{2012ApJ...747..144M} illustrated a putative
template for what a potential planetary system around Proxima could resemble.
Notably, the planets were all sub-Earth sized, and would thus have eluded the
radial velocity search efforts of \citet{2008A+A...488.1149E} and
\citet{2009A+A...505..859Z}, should similar planets orbit Proxima. More over,
the planets were at extreme proximity to the star, with periods ranging from
0.45\,d to 1.86\,d, leading to sizable geometric transit probabilities. Indeed,
should Proxima harbor a Kepler-42 like system, the transit probability would be
$\sim$10\%.

Second, \Kepler\ occurrence rate statistics showed that planets around early
M-dwarfs are very common, with an average of $(2.5\pm0.2)$ planets per star
\citep{2015ApJ...807...45D}. Radial velocity campaigns come to similar
conclusions, finding evidence for at least one planet per M-dwarf
\citep{2014MNRAS.441.1545T}. Together then, this implies that Proxima not only
has an excellent chance of harboring a planetary system but such planets have
a reasonable probability of transiting and producing mmag level signals. These
arguments inspired our team to conduct a transit survey of Proxima starting in
2014 with the \textit{Microwave and Oscillations of Stars} (hereafter \MOST)
telescope.

\MOST\ is a 53\,kg satellite in Low Earth Orbit (LEO) with a 15\,cm aperture
visible band camera (35-750\,nm). \MOST\ is able to deliver mmag level 
photometry (for $V\lesssim12$) at high cadence over several weeks baselines,
although observations are typically interrupted once per 101\,minute orbit as
the spacecraft passes behind the Earth. \MOST\ has been successful in 
discovering several new transiting systems, such as 55~Cnc~e 
\citep{2011ApJ...737L..18W}, HD~97658~b \citep{2013ApJ...772L...2D} and most 
recently HIP~116454~b \citep{2015ApJ...800...59V}. For Proxima Centauri, we 
estimated \MOST\ should deliver $\sim0.3$\,mmag precision photometry on an hour
timescale, making it well suited for detecting the 4.2\,mmag transit expected 
to be caused by an Earth-sized planet and thus two seasons of observations were
undertaken in 2014 and 2015.

Evidently, our team was not alone in returning to Proxima, with the
\textit{Pale Red Dot} campaign (PRD hereafter) conducting their own intensive
search using radial velocities in 2016. By combining the PRD data with previous
radial velocities, \citet{nature} recently announced the detection of a 11.2\,d
planetary candidate- Proxima b. Since radial velocities do not reveal the 
inclination of the planetary orbit, only the minimum mass of Proxima b is 
presently known at $M_P \sin i = 1.27_{-0.17}^{+0.19}$\,\mearth. Since the 
transition from Terran (solid-like) to Neptunian worlds occurs at 
$(2.1\pm0.6)$\,\mearth\ \citep{chen}, the compositional nature of Proxima b is 
presently ambiguous. If transits of Proxima b were observed, the inclination 
could be resolved, as well as offering the opportunity to further characterize 
this remarkable world.

In this work, we present the results of our search for transiting planets around
Proxima Centauri with \MOST\ photometry. We describe the observations and data
treatment stages in Section~\ref{sec:data} and our photometric model in 
Section~\ref{sec:model}. In Section~\ref{sec:results}, we present the results of a
localized search using the reported Proxima b ephemeris, followed by two sets of
tests in Sections~\ref{sec:tests}\&\ref{sec:hats}. Finally, we discuss the
constraints our data place on Proxima b in Section~\ref{sec:discussion}.

\clearpage
\section{OBSERVATIONS}
\label{sec:data}

\subsection{\MOST\ Observations}

\MOST\ observed Proxima Centauri in May 2014 (beginning on HJD(2000) 
2456793.18) for about 12.5 days. Proxima Centauri falls outside of the 
Continuous Viewing Zone of \MOST\ ($-19^{\deg}$ to $+36^{\deg}$ in declination;
see \citealt{2003PASP..115.1023W}) and can only be observed for a fraction of 
the satellite’s 101 min orbit. For this reason, and for other science queue 
considerations, data was collected for about 30\% of each \MOST\ orbit and was 
sampled at an average rate of 63.4 seconds. \MOST\ again observed Proxima 
Centauri in May 2015 (starting on HJD(2000) 2457148.54), this time for a total 
of 31 days with extended coverage to almost 50\% of every \MOST\ orbit. Data 
were again sampled at an average rate of about 63 seconds.  

Flux measurements were extracted from each image using aperture photometry 
techniques outlined by \citet{2006MmSAI..77..282R}. Background counts, 
inter-pixel correlations, and pointing drifts were accounted for by 
subtracting polynomials fitted through correlations between the measured target
flux and those parameters. Removal of stray Earth-shine onto the CCD was done 
by folding the time series at the orbital period of MOST and subtracting a 
running mean through 30 orbital bins. Any remaining statistical outliers were 
removed, resulting in $\sim$2600 individual time series measurements from the
2014 data set and $\sim$13000 data points from 2015. 

\begin{figure*}
\begin{center}
\includegraphics[width=17.0cm,angle=0,clip=true]{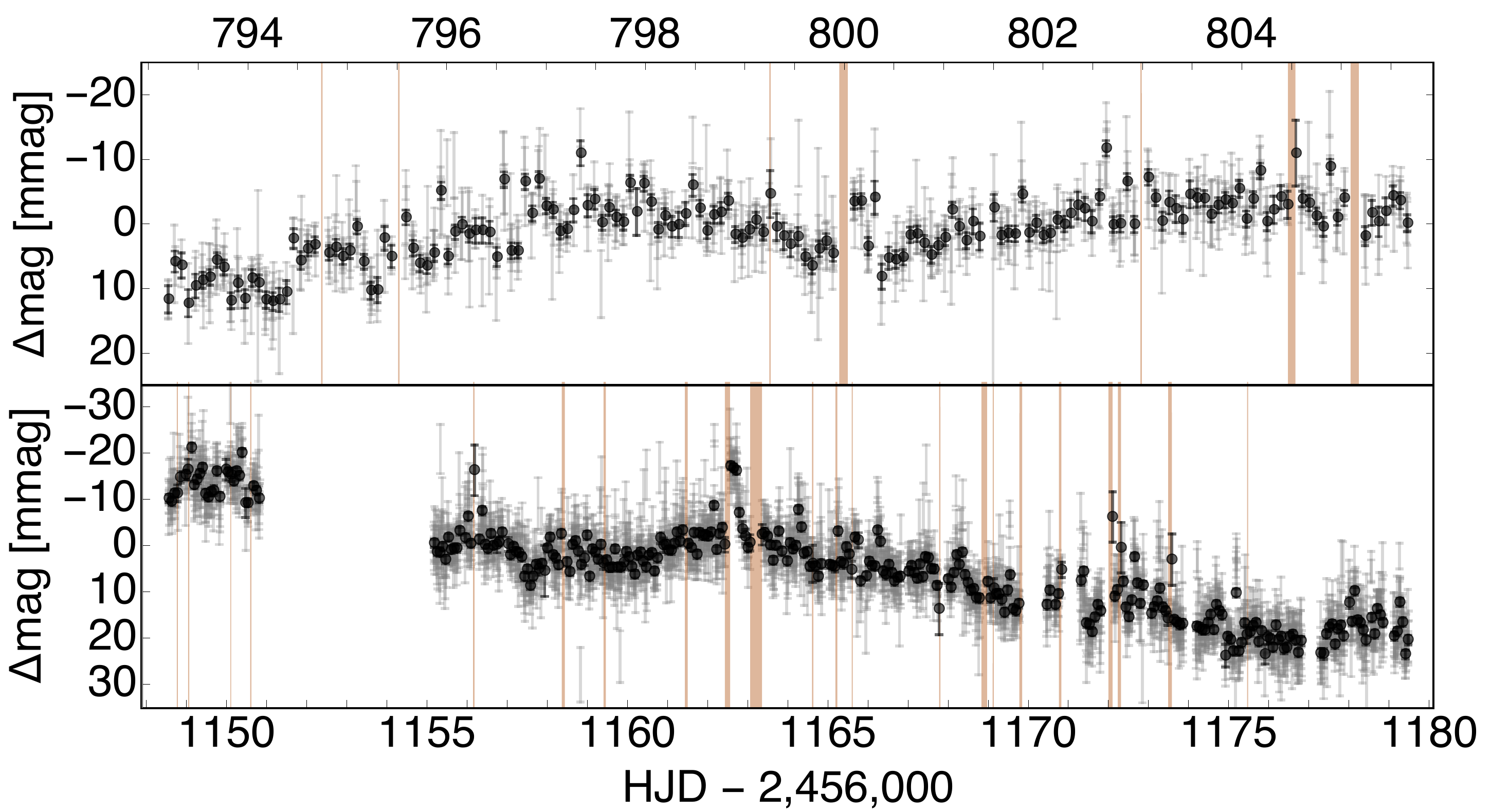}
\caption{
Corrected photometric observations of Proxima with \MOST\ in 2014 (top)
and 2015 (bottom) shown with 5$\times$cadence binning ($\sim5.3$\,minutes).
Each orbital visit of \MOST\ is binned together in black. The vertical colored
regions denote regions ignored in this work, since flares were identified by
\citet{davenport}.
}
\label{fig:data}
\end{center}
\end{figure*}

The time series was then inspected for flare-like events using
v1.3.11 of the flare-finding suite \FBEYE\ from \citet{2014ApJ...797..122D}.
The results of this exercise are discussed in detail in the accompanying
paper of \citet{davenport}, but for the purposes of this work these points
are removed in all subsequent analyses of the photometry. The location of
these events are highlighted in Figure~\ref{fig:data}.

\subsection{Time-Correlated Structure and Trends}
\label{sub:trends}

After correcting the photometry and removing the flares, it is clear
that our \MOST\ data exhibits time-correlated structure in both
seasons (see Figure~\ref{fig:data} and Table~\ref{tab:datatable}). Long-term
trends are pronounced in both seasons, with 2014 displaying a slow
brightness increase along with a sinusoidal-like few mmag variation
on the timescale of a week. In 2015, the structure appears more complex
and exhibits a slow brightness decrease. These trends are not seen in
any of the comparison stars and thus we identify them as being
astrophysical in nature.

The slow brightness trends may be associated with the claimed 83\,d
rotation period of Proxima Centauri \citep{1998AJ....116..429B}. The
remaining, and quite pronounced, structure may be a result of frequent
flaring \citep{davenport} and associated corononal mass ejections, as
well as magnetic activity such as evolving spots, plages and networks.

Interpreting the origin of the observed structure is beyond the scope of this
work, for which this structure represents a impediment in our ability
to detect putative transits of Proxima.

\begin{table}
\caption{Reduced \MOST\ photometry used in this work, excluding times afflicted by large flares.
A portion of the table is shown here, the full version is available in the electronic version
of the paper and at \href{\urllink}{this URL}.
} 
\centering 
\begin{tabular}{c c c} 
\hline\hline
HJD$_{\mathrm{UTC}}$ - 2,451,545 & $\Delta$mag & Uncertainty \\ [0.5ex] 
\hline
5248.197851471566	& 0.0118 & 0.0030 \\
5248.200048126498	& 0.0116 & 0.0030 \\
5248.259628134210	& 0.0028 & 0.0026 \\
5248.262194952687	& 0.0092 & 0.0030 \\
5248.270990357013	& 0.0065 & 0.0030 \\
5248.329765768074	& 0.0072 & 0.0026 \\
5248.332326375948	& 0.0054 & 0.0030 \\
5248.399228360851	& 0.0155 & 0.0030 \\
\vdots & \vdots & \vdots \\
\hline 
\end{tabular}
\label{tab:datatable} 
\end{table}

\subsection{Gaussian Process (GP) Regression}
\label{sub:GP}

In order to search for transits, the structure and trends present in our
data require modeling. For reasons described in what follows, we elected
to use Gaussian Process (GP) regression to model out this structure. Here,
one assumes the data is distributed around the transit model as a multivariate
Gaussian including off-diagonal elements within its covariance matrix, 
$\boldsymbol{\Sigma}$. This eliminates the assumption of independent 
uncertainties and allows each point in the time series to have some degree of
correlation with every other point. The log-likelihood function, used for
subsequent regression, may be written as

\begin{align}
\log\mathcal{L} &= - \tfrac{1}{2} \mathbf{r}^{T} \boldsymbol{\Sigma}^{-1} \mathbf{r}
                   - \tfrac{1}{2} \log \mathrm{det} \boldsymbol{\Sigma}
									 - \tfrac{n}{2}\log2\pi,
\label{eqn:loglike}
\end{align}

where $\mathbf{r}$ is a vector of the residuals between the transit model and
the data. The very large number of covariance matrix elements are a-priori 
unknown to us, but GPs model the covariance matrix with some assumed smooth,
functional form, known as the kernel. The kernel, $\mathbf{K}$, is described by one or more
hyper-parameters, $\boldsymbol{\theta}_{\mathrm{hyper}}$, which are freely 
explored along with the usual model parameters,
$\boldsymbol{\theta}_{\mathrm{transit}}$, during the fitting procedure.

GPs have emerged as one of the most popular and successful methods of modeling
time correlated noise in the analysis of transit photometry 
\citep{2012MNRAS.419.2683G,2015MNRAS.451..680E,2015Natur.527..204B} and are
appealing for their ability to model complex structure with relatively few
new regression parameters. However, inverting the covariance matrix at
each realization is computationally expensive and typically GPs are
computationally prohibitive for $n\gtrsim\mathcal{O}[10^3]$ data points,
which is particularly relevant in this work given that we have over $10^4$
photometric measurements.

\subsection{Binning and Kernel Selection}
\label{sub:binning}

To overcome the computational challenge of inverting the covariance matrix,
one may first apply modest binning to the time series. Ideally, the relevant
correlation time scale(s) should be significantly greater than the time scale
used for binning, such that correlations are preserved.

The native cadence of our photometric measurements is 63.5\,seconds and after
removing outliers and flares, we have 2461 data points in the 2014 season
and 11473 points in 2015. We first assume the kernel parameters for each season
are wholly independent, given the large change in time. We then focus on the
null model of a transit-free case, where the data is solely described by an
offset parameter, $a$, and the GP. Accordingly, the 2014 season can be treated
independent of 2015.

We found that 2461 data points was not a computationally prohibitive number of
points for GP regression, which allows us to directly compare the inferred
GP kernel parameters between the binned and unbinned data. We set a binning
time scale equal to 317.5\,seconds (equivalent to 5 consecutive cadences), or
approximately five minutes, and employ temporal windows for the binning rather
than $N$-point binning, due to the considerable number of data gaps present.
For the GP kernel, we adopted the popular Mat\'ern-3/2 kernel
given by

\begin{align}
\mathbf{K}_{i,j}(\alpha,l) &= \alpha^2 \Bigg( 1 + \frac{\sqrt{3}|t_i-t_j|}{l} \Bigg) \exp\Bigg( -\frac{\sqrt{3}|t_i-t_j|}{l} \Bigg),
\label{eqn:M32}
\end{align}

where $l$ controls the time scale of correlations and $\alpha$ controls
the magnitude. 
The full covariance matrix, $\boldsymbol{\Sigma}$, is the sum of the
matrix $\mathbf{K}$ and a diagonal matrix of the square of the measurement uncertainties.
We then regressed both versions of the 2014 time series
using \multi\ \citep{2008MNRAS.384..449F,2009MNRAS.398.1601F} and computed
parameter posteriors. The agreement between the two is excellent, with
$a=0.79_{-0.57}^{+0.59}$\,mmag from the unbinned data versus
$a=0.78_{-0.58}^{+0.58}$\,mmag using the binned. Similarly, the $l$
correlation time scale is almost identical- $l=130_{-15}^{+16}$\,mins
in the unbinned versus $l=130_{-15}^{+17}$\,mins in the binned. Further,
this time scale is much greater than the 5\,minute binning time scale
adopted, ensuring key correlations are not affected by the binning procedure.

For the 2015 data, we are unable to repeat this test given the much larger
number of unbinned points. However, regressing the same model and GP on the
2015 binned data reveals $l=269_{-15}^{+15}$\,mins, which is again much greater
than the binning time scale (in fact even more so than before). Accordingly,
we conclude that the 5\,minute binning procedure does not affect the GP
inference nor should affect our ability to detect transits, given that such
events occur on significantly longer time scales too. Our final binned time
series includes 709 points in the 2014 season and 2850 points in 2015, which
are the points plotted in gray in Figure~\ref{fig:data}.

\begin{figure*}
\begin{center}
\includegraphics[width=17.0cm,angle=0,clip=true]{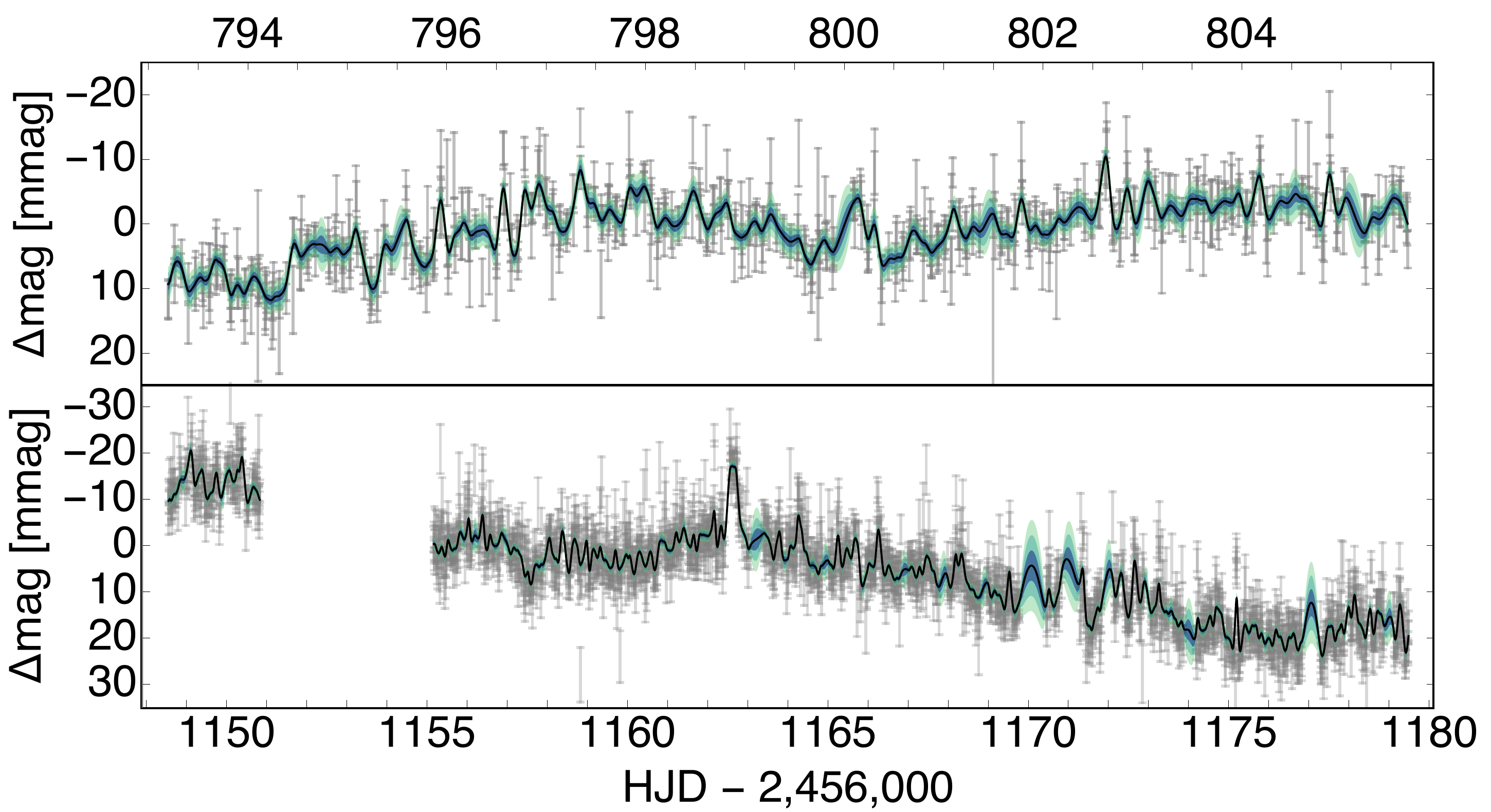}
\caption{
Corrected photometric observations of Proxima with \MOST\ in 2014
(top) and 2015 (bottom) shown with 5$\times$cadence binning 
($\sim5.3$\,minutes) with large flares excluded. The black line is
the maximum likelihood fit of a GP using a Mat\'ern-3/2 kernel, with
the 0.5, 1.0 and 1.5\,$\sigma$ confidence regions denoted by the
gray regions (GP shown is from model $\mathcal{M}_2$).
}
\label{fig:null}
\end{center}
\end{figure*}

Although we assumed a Mat\'ern-3/2 kernel in these tests, several other
commonly used kernels are investigated before continuing. We compared the
Bayesian evidence (or marginal likelihood) resulting from re-fitting both
seasons of data for a Mat\'ern-3/2 kernel, Mat\'ern-3/2 quasi-periodic
kernel, a Mat\'ern-5/2 kernel and a squared-exponential kernel. After
conducting these fits on the 2014 unbinned, 2014 binned and 2015 binned data,
with identical priors, we find in all cases that the Mat\'ern-3/2 kernel
is favored, as shown in Table~\ref{tab:GPtests}. We therefore adopt the
Mat\'ern-3/2 kernel in all subsequent photometric analysis of Proxima Centauri
and show this favored GP over-plotted with the data in 
Figure~\ref{fig:null}.

\begin{table}
\caption{Comparison of Bayesian evidences, $\mathcal{Z}$, for four commonly 
used kernel choices. The preferred model for each data set is emboldened.} 
\centering 
\begin{tabular}{l c} 
\hline
Kernel & $\log\mathcal{Z}$ \\ [0.5ex] 
\hline
\emph{2014 unbinned} \\
\hline
Squared-Exponential & $9856.304 \pm 0.088$ \\
\textbf{Mat\'ern-3/2} & $\mathbf{9865.807 \pm 0.083}$ \\
Mat\'ern-5/2 & $9862.166 \pm 0.085$ \\
Quasi-periodic Mat\'ern-3/2 & $9865.513 \pm 0.084$ \\
\hline
\emph{2014 binned} \\
\hline
Squared-Exponential & $2989.686 \pm 0.088$ \\
\textbf{Mat\'ern-3/2} & $\mathbf{2998.977 \pm 0.084}$ \\
Mat\'ern-5/2 & $2995.398 \pm 0.085$ \\
Quasi-periodic Mat\'ern-3/2  & $2998.795 \pm 0.084$ \\
\hline
\emph{2015 binned} \\
\hline
Squared-Exponential & $11192.194 \pm 0.109$ \\
\textbf{Mat\'ern-3/2} & $\mathbf{11337.855 \pm 0.100}$ \\
Mat\'ern-5/2 & $11284.873 \pm 0.102$ \\
Quasi-periodic Mat\'ern-3/2 & $11337.101 \pm 0.103$ \\ [1ex]
\hline\hline 
\end{tabular}
\label{tab:GPtests} 
\end{table}

\section{PRIORS, MODELS \& TESTS}
\label{sec:model}

\subsection{Predicting the Transit Ephemeris}
\label{sub:ephemeris}

The radial velocity solution of \citet{nature} provides joint posterior
distributions for numerous parameters including the orbital period, $P$, and
the mean anomaly at a reference time $t_0$, $M_0$. We first converted the $M_0$
column into a time of inferior conjunction, $t_{\mathrm{IC}}$, via Kepler's
equation. For low-eccentricity orbits, such as that of Proxima b ($e<0.3$ to 
95\% confidence; \citealt{nature}), the time of transit minimum, $\tau$, is
equal to the time of inferior conjunction \citep{thesis}.

The time of inferior conjunction can be computed at any epoch of our choosing by
adding on some integer number of periods. We therefore elected to calculate 
$t_{\mathrm{IC}}$ at every possible epoch from $-200$ to $+200$ orbital periods.
For each realization, we computed the standard deviation of the resulting posterior
and also the correlation with respect to the $P$ posterior samples. We found that
both are minimized for $-65^{\mathrm{th}}$ epoch, for which
$t_{\mathrm{IC}} = 2,456,678.78\pm0.56$\,HJD. Both this term, and the orbital period
of $P = (11.1856\pm0.0013)$\,d, are well approximated as two independent normal
distributions.

\subsection{Predicting the Radius}
\label{sub:radius}

In order to guide our targeted search for transits of Proxima b, we first
estimate the amplitude of the transit signal expected. If the mass of a planet
is known, the radius can be predicted using an empirical mass-radius relation.
In this work, we use the relation of \citet{chen} which is probabilistic,
includes freely inferred transitional regions and was calibrated on the widest
range of data available.

If the planet is transiting, this imposes the condition that $b<1+p$. Given that
Proxima b is an Earth-mass planet, we expect $p \simeq 0.06$ and thus $p\ll1$.
This allows us to write that a transiting Proxima b must satisfy

\begin{align}
\cos i < \frac{1}{a_R} \frac{1-e^2}{1+e\sin\omega}.
\end{align}

Using the joint posterior distribution samples for $P$, $e$ and $\omega$ from
\citet{nature}, this condition requires $i<89.1^{\circ}$ to 95.45\% confidence.
For any transiting planet then, the $\sin i$ effect on the true mass is much
smaller than the present measurement uncertainty on $M_P \sin i$. Accordingly,
we may simply adopt $M_P \simeq M_P\sin i$ in estimating the radius of a
transiting Proxima b.

We now use the posterior samples of $M_P \sin i$ to estimate a probabilistic
range for $R_P$ using the \forecaster\ code of \citet{chen}. This estimate
accounts for the measurement uncertainty of \citet{nature}, the measurement
uncertainties in the calibration of \citet{chen} and the intrinsic dispersion
in radii observed as a function of mass \citep{chen}. Under the assumption that
Proxima b is transiting, we estimate that 
$R_P = 1.06_{-0.11}^{+0.13}$\,$R_{\oplus}$. Normalizing by the radius of the
star \citep{radius}, we predict $p=R_P/R_{\star}=0.0693_{-0.0083}^{+0.0095}$,
which is well fit by a logistic distribution with shape parameters
$\mu=0.069$ and $s=0.0051$. We also find that 99\% of the posterior samples
for $p$ satisfy $p<0.1$ and conservatively double this limit to $p<0.2$ as
a truncation point to our prior. This also provides a cut-off for the impact
parameter of $b<1.2$.

\subsection{Models Considered and Associated Priors}
\label{sub:priors}

We considered three different transit models in our targeted search for
transits of Proxima b, where we varied the degree of prior information we used
from the \citet{nature} discovery. We label the models as $\mathcal{M}_1$, 
$\mathcal{M}_2$ and $\mathcal{M}_3$, where the subscript increases with 
increased use of prior information. These fits may be compared directly to the
null fit of $\mathcal{M}_0$, where no transit is included and only the GP
hyper-parameters, $\boldsymbol{\theta}_{\mathrm{hyper}}$, are fitted.

We list the priors used in Table~\ref{tab:priors}, where it can be seen that
the GP hyper-priors are identical in all fits. This allows us to compare the
Bayesian evidences between each model to aid model selection. All three
transit models used the normal prior on orbital period, otherwise the search
would be blind rather than targeted. $\mathcal{M}_2$ uses an informative
prior on the time of transit minimum, unlike $\mathcal{M}_1$. The final model,
$\mathcal{M}_3$, also uses these period and phase informative priors plus 
an informative prior on the radius of the planet, as computed earlier in
Section~\ref{sub:radius}.

\begin{table*}
\caption{
Priors used in the targeted transit search of Proxima b, spanning four different models
(each column). $\mathcal{U}$ denotes a uniform prior, $\mathcal{W}$ a wrap-around
uniform, $\mathcal{J}$ a log-uniform, $\delta$ a Delta function prior and $\mathcal{X}$
a logistical distribution prior. We set the reference time to
$t_{\mathrm{ref}}=2457165.385$\,HJD.
}
\centering
\begin{tabular}{l c c c c}
\hline\hline
Parameter & $\mathcal{M}_0$ & $\mathcal{M}_1$ & $\mathcal{M}_2$ & $\mathcal{M}_3$ \\ [0.5ex]
\hline
$a_{\mathrm{2014}}$ [mmag]	& $\mathcal{U}[-10^{-2},10^{-2}]$ & $\mathcal{U}[-10^{-2},10^{-2}]$ & $\mathcal{U}[-10^{-2},10^{-2}]$ & $\mathcal{U}[-10^{-2},10^{-2}]$ \\
$\alpha_{\mathrm{2014}}$		& $\mathcal{J}[-10^{-1},10^{1}]$ & $\mathcal{J}[-10^{-1},10^{1}]$ & $\mathcal{J}[-10^{-1},10^{1}]$ & $\mathcal{J}[-10^{-1},10^{1}]$ \\
$l_{\mathrm{2014}}$					& $\mathcal{J}[-10^{-2},10^{0}]$ & $\mathcal{J}[-10^{-2},10^{0}]$ & $\mathcal{J}[-10^{-2},10^{0}]$ & $\mathcal{J}[-10^{-2},10^{0}]$ \\
$a_{\mathrm{2015}}$ [mmag]	& $\mathcal{U}[-10^{-2},10^{-2}]$ & $\mathcal{U}[-10^{-2},10^{-2}]$ & $\mathcal{U}[-10^{-2},10^{-2}]$ & $\mathcal{U}[-10^{-2},10^{-2}]$ \\
$\alpha_{\mathrm{2015}}$		& $\mathcal{J}[-10^{-1},10^{1}]$ & $\mathcal{J}[-10^{-1},10^{1}]$ & $\mathcal{J}[-10^{-1},10^{1}]$ & $\mathcal{J}[-10^{-1},10^{1}]$ \\
$l_{\mathrm{2015}}$					& $\mathcal{J}[-10^{-2},10^{0}]$ & $\mathcal{J}[-10^{-2},10^{0}]$ & $\mathcal{J}[-10^{-2},10^{0}]$ & $\mathcal{J}[-10^{-2},10^{0}]$ \\
\hline
$p=(R_P/R_{\star})$ & - & $\mathcal{J}[10^{-3},10^{-0.70}]$ & $\mathcal{J}[10^{-3},10^{-0.70}]$ & $\mathcal{X}[0.069,0.0050]$ \\ 
$b$ & - & $\mathcal{U}[0,1.2]$	& $\mathcal{U}[0,1.2]$ & $\mathcal{U}[0,1.2]$ \\
$\tau$ [HJD-2,456,000] & - & $\mathcal{W}[t_{\mathrm{ref}},t_{\mathrm{ref}}+P]$ & $\mathcal{N}[678.78,0.59]$ & $\mathcal{N}[678.78,0.59]$\\
$P$\,[days] & - & $\mathcal{N}[11.1856,0.0013]$	& $\mathcal{N}[11.1856,0.0013]$ & $\mathcal{N}[11.1856,0.0013]$ \\
$\rho_{\star}$\,[kg\,m$^{-3}$] & - & $\delta[10^{4.792}]$ & $\delta[10^{4.792}]$ & $\delta[10^{4.792}]$ \\
$e$ & - & $\delta[0]$ & $\delta[0]$ & $\delta[0]$ \\
\hline
$\log \mathcal{Z}$ - 14,300	& $36.763\pm0.065$ & $46.286\pm0.089$ & $38.290\pm0.079$ & $41.336\pm0.077$ \\ [1ex]
\hline
\end{tabular}
\label{tab:priors}
\end{table*}

Limb darkened transits are generated using the \citet{mandel} algorithm.
Two quadratic limb darkening coefficients are kept fixed at $u_1=0.7948$ and
$u_2=0.0825$, estimated by finding nearest neighbor interpolation of the 
PHOENIX model grids for \MOST\ generated in \citet{2014A+A...567A...3C}
(using $\log g=5.25$ and $T_{\mathrm{eff}}=3050$\,K). Eccentricity is kept 
fixed at zero, since Proxima b has a low eccentricity \citep{nature}. 
Similarly, we fix the mean stellar density of the star, which is very well 
constrained given that Proxima is one of the most well-studied M-dwarfs. 
These fixed parameters significantly reduce the number of parameters to 
explore, making the calculation of the Bayesian evidence of models using 
Gaussian processes with several thousand data points computationally 
feasible. By fixing these terms, transit parameter inferences may slightly
underestimate the true uncertainties but since our primary objective is 
signal detection, the ability to be able to feasibly compute evidences 
outweighs this cost, in our view.

\subsection{Mis-specified Likelihood Function}
\label{sub:badlike}

As discussed earlier, the likelihood function used in this work is that of
a Gaussian process, described in Section~\ref{sub:GP}, and given by 
Equation~\ref{eqn:loglike}. Additionally, we are computing evidences using
\multi, in order to conduct model comparison. Note that the compuational
expense of this work prohibited us from computing evidences with several
different methods, and thus we adopt those from \multi\ only in what follows.

A common perception of GPs is that they are extremely flexible models, 
seemingly able to model out just about any observed correlated noise structure,
particularly when one regresses the GP kernel hyper-parameters simultaneously 
with the model. Indeed, \citet{2016MNRAS.461.2440F} go as far as to actively 
caution against using GPs since they lead to frequently missing true signals
due their over-zealous ability to fit out time series structure. This logic 
suggests that a GP-only model ($\mathcal{M}_0$) would generally be favored over
a GP+transit model ($\mathcal{M}_2$ \& $\mathcal{M}_3$) even when real signals 
are present. Or, equivalently, it implies that our evidences may be 
conservative and the actual weight of evidence for the transit models may be 
higher than that formally calculated.

This logic can be flawed if our likelihood function is mis-specified, which 
means that the marginal likelihood would be inaccurate. This can occur if the
assumed functional form of the GP kernel is a poor approximation of the true
(and unknown) covariance matrix. In this study, the high flare activity of 
Proxima makes this a plausible scenario. By extrapolating the rates of large
flares, \citet{davenport} estimate that Proxima exhibits a 0.5\% brightness 
increase once every $\sim$20\,minutes (on average). Note that this issue is not
limited to just \MOST\ data but implies that any visible light photometry of 
Proxima will be affected by ostensibly constant stochastic deviations at the
level of 5\,mmag, as a result of flares. A Mat\'ern 3/2 kernel is not designed
to describe a superposition of flare events and thus formally we expect our
likelihood function to be mis-specified for this reason.

In conclusion, the marginal likelihoods from our fits will not be accurate.
It is not clear what alternative kernel or likelihood function could deal
with this kind of noise structure either, and thus we still favor GPs over any
alternatives. Although our evidences will be inaccurate, they may still be
useful in guiding which models are preferred. Even if the Bayes factor is
inaccurate, this does not mean it cannot be used to rank models in order of
preference, since, after all, the majority of the residuals are indeed normally
distributed. However, any candidate solutions from this process must be
treated with great caution and subject to higher scrutiny and skepticism than
usual.

\section{RESULTS}
\label{sec:results}

\subsection{Signal S}
\label{sub:signalS}

We first discuss the results of model $\mathcal{M}_1$, where the transit
phase is described by an uninformative prior. The marginal likelihood
indicates a strong preference for $\mathcal{M}_1$ over $\mathcal{M}_0$,
with $\Delta\log\mathcal{Z}=9.52\pm0.11$. Hereafter, we refer to this
solution as signal S.

\begin{figure*}
\begin{center}
\includegraphics[height=5.75cm,angle=0,clip=true]{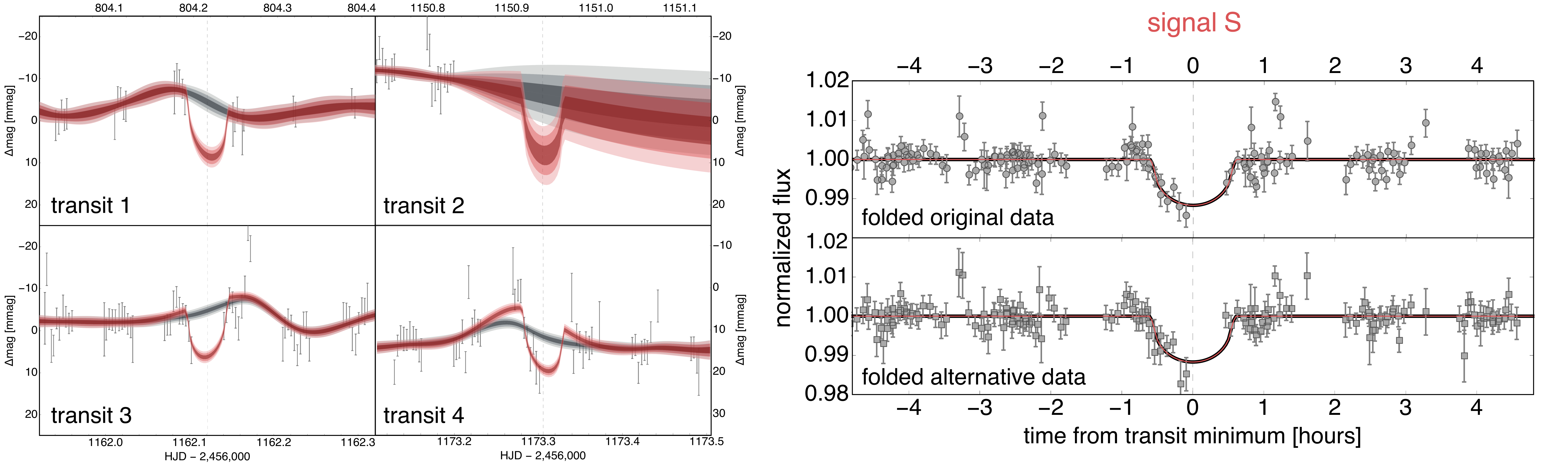}
\caption{
\textbf{Left:} Zoom-in of the Proxima b transits for the spurious signal
(signal S) from models $\mathcal{M}_1$. The GP+transit model 0.5,
1.0 and 1.5\,$\sigma$ confidence regions are shown by the colored regions,
whereas the GP-only (model $\mathcal{M}_0$) is depicted in gray.
\textbf{Right:} Phase folded light curve of the transit signal S after 
removing the GP component of model $\mathcal{M}_1$ using the nominal
\MOST\ data (upper panel) and an independent re-processing (lower panel).
}
\label{fig:signalS}
\end{center}
\end{figure*}

The $\mathcal{M}_1$ ephemeris yields four transit epochs within our
\MOST\ time series, although one of these occurs during a data gap, as
shown in Figure~\ref{fig:signalS}. We note that signal S is primarily 
driven by a large feature in the fourth epoch, at HJD\,$2457173.3$. To
ensure that the recovered signal was not an artifact of our data processing
method, one of us (J. Rowe) re-processed the data independently. As shown
by the square points in Figure~\ref{fig:signalS}, the signal appears coherent
in both data products.

The time of transit minimum, $\tau$, has a non-Gaussian but narrow marginal
posterior with a 1\,$\sigma$ credible interval of 
$\tau=1150.9348_{-0.0026}^{+0.0019}$\,(HJD-2,456,000), which deviates 
substantially from the predicted time based on the radial velocity fits of
\citet{nature}. For reference, all of the model parameter credible intervals,
from all four models, are listed in Table~\ref{tab:Mparams}.

\begin{table*}
\caption{A-posteriori median and 68.3\% credible intervals of each model parameter
for the four models regressed to the \MOST\ photometry. Full posterior samples
are available at \href{\urllink}{this URL}.
$^{\dagger}$ = assuming a fixed stellar radius of $0.123$\,$R_{\odot}$.
} 
\centering 
\begin{tabular}{c c c c c} 
\hline\hline
Parameter & $\mathcal{M}_0$ & $\mathcal{M}_1$ & $\mathcal{M}_2$ & $\mathcal{M}_3$ \\ [0.5ex] 
\hline
$a_{\mathrm{2014}}$\,[mmag] & $0.78_{-0.59}^{-0.59}$ & $0.78_{-0.54}^{+0.55}$ & $0.77_{-0.57}^{+0.57}$	& $0.78_{-0.58}^{+0.58}$ \\
$\alpha_{\mathrm{2014}}$    & $1.44_{-0.10}^{+0.12}$ & $1.44_{-0.09}^{+0.10}$ & $1.44_{-0.10}^{+0.11}$	& $1.45_{-0.10}^{+0.11}$ \\
$l_{\mathrm{2014}}$\,[mins] & $131_{-15}^{+16}$      & $131_{-14}^{+15}$      & $130_{-14}^{+15}$	    & $130_{-15}^{+16}$      \\
$a_{\mathrm{2015}}$\,[mmag] & $5.41_{-0.87}^{+0.87}$ & $5.38_{-0.80}^{+0.82}$ & $5.40_{-0.85}^{+0.86}$	& $5.40_{-0.89}^{+0.89}$ \\
$\alpha_{\mathrm{2015}}$    & $2.30_{-0.10}^{+0.11}$ & $2.30_{-0.09}^{+0.11}$ & $2.34_{-0.10}^{+0.11}$	& $2.34_{-0.10}^{+0.11}$ \\
$l_{\mathrm{2015}}$\,[mins] & $269_{-14}^{+16}$      & $268_{-13}^{+15}$      & $274_{-14}^{+16}$	    & $274_{-15}^{+16}$      \\
\hline
$R_P$\,$[R_{\oplus}]^{\dagger}$ & - & $1.38_{-0.12}^{+0.11}$ & $1.23_{-0.15}^{+0.13}$ & $1.14_{-0.10}^{+0.10}$ \\
$b$                           & - & $0.22_{-0.14}^{+0.19}$ & $0.28_{-0.19}^{+0.24}$ & $0.25_{-0.18}^{+0.24}$ \\
$\tau$ [HJD-2,456,000]        & - & $983.1656_{-0.0330}^{+0.0064}$   & $980.0554_{-0.0023}^{+0.0027}$   & $980.0552_{-0.0026}^{+0.0029}$   \\
$P$\,[days]                   & - & $11.18467_{-0.00039}^{+0.00200}$ & $11.18725_{-0.00016}^{+0.00012}$ & $11.18723_{-0.00019}^{+0.00014}$ \\
\hline 
\end{tabular}
\label{tab:Mparams} 
\end{table*}

To quantify the ephemeris disagreement, we used the posterior samples of the 
radial velocity predicted time of transit minimum, computed earlier in 
Section~\ref{sub:ephemeris}, and propagated the joint posterior of $P$ and 
$\tau$ to the equivalent epoch, giving 
$\tau=1148.59_{-0.59}^{+0.59}$\,(HJD-2,456,000). The $p$-value of signal S's 
$\tau$ posterior exceeds 4\,$\sigma$ and is difficult to reconcile with radial
velocity solution. Note that the radial velocity posteriors include a floating
eccentricity and thus this effect is accounted for here. This is point is 
formally established in the Bayesian framework by the fact that 
models $\mathcal{M}_2$ and $\mathcal{M}_3$
do not recover signal S when using the radial velocity derived $\tau$ prior.

We also note that the inferred planetary radius is at the 2\,$\sigma$ upper
limit of the \forecaster\ prediction, which although not concerning in
isolation does compound upon these earlier concerns. Further, visual inspection
of the transits (Figure~\ref{fig:signalS}) shows that the signal is primarily
driven by an apparent flux increase around the times of transit, rather than
a flux decrease, which raises additional skepticism.

Whilst one could, in principle, refit the radial velocities imposing this 
transit phase as a prior, that model would be implicitly assuming that signal S
is real - an assumption which is not warranted given the challenging noise 
structure of our data set and the arguments made above. The incompatibility of 
the transit phase, and to a lesser degree the poor phase coverage and inflated 
radius, lead us to conclude that signal S is unlikely associated with
Proxima b and is either spurious due to flare-induced likelihood 
mis-specification or potentially an additional transiting planet, driven by
a single event within our data.

\subsection{Signal C}
\label{sub:signalC}

We next consider the results of models $\mathcal{M}_2$ and $\mathcal{M}_3$,
both of which recover the same transit signal, which we hereafter refer to as
signal C. Three modes are recovered by the fit of $\mathcal{M}_2$, but the 
dominant mode is strongly favored at a Bayes factor of 20.1. Of the three 
modes, only the dominant is preferred over the null model of $\mathcal{M}_0$ 
and it is this mode which is compatible with the signal recovered by model 
$\mathcal{M}_3$. We therefore ignore the other two modes in what follows.

Whilst the two models recover the same signal, $\mathcal{M}_2$ is favored over
the null hypothesis with a Bayes factor of 4.6 whereas $\mathcal{M}_3$ is much
stronger at 96.8. This can be understood by the fact the two models recover 
very similar $p$ posteriors 
($R_P(\mathcal{M}_2)=1.23_{-0.15}^{+0.13}$\,$R_{\oplus}$ versus
$R_P(\mathcal{M}_2)=1.14_{-0.10}^{+0.10}$\,$R_{\oplus}$) but $\mathcal{M}_2$
used a broad, uninformative prior over which the average likelihood
is lower, thus penalizing the model for effectively being more complicated.

\begin{figure*}
\begin{center}
\includegraphics[height=5.75cm,angle=0,clip=true]{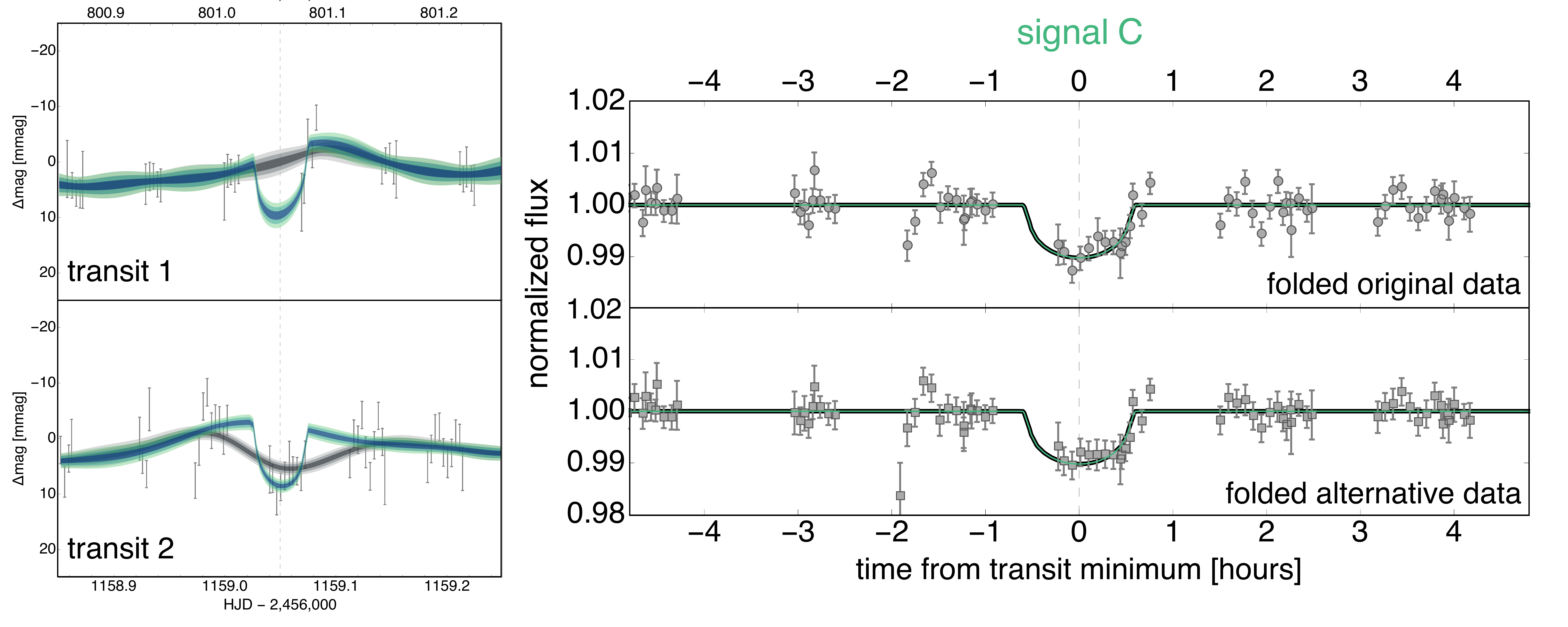}
\caption{
Same as Figure~\ref{fig:signalS} except for the candidate transit signal of
Proxima b, signal C. Right panels reproduce the left panels, except including
the independently processed \MOST\ data in lilac for comparison.
}
\label{fig:signalC}
\end{center}
\end{figure*}

The signal is shown in Figure~\ref{fig:signalC} where we highlight how once
again the independent re-processing of the \MOST\ data produces a consistent
signal.

The phase from both models is not incompatible with the radial velocity 
constraints, giving a $p$-value of 1.56\,$\sigma$, although this is to be 
expected since it was imposed as an informative prior. A more useful test is
that the freely fitted radius from model $\mathcal{M}_2$ is compatible with
the radius prediction from \forecaster.

\begin{figure*}
\begin{center}
\includegraphics[width=17.0cm,angle=0,clip=true]{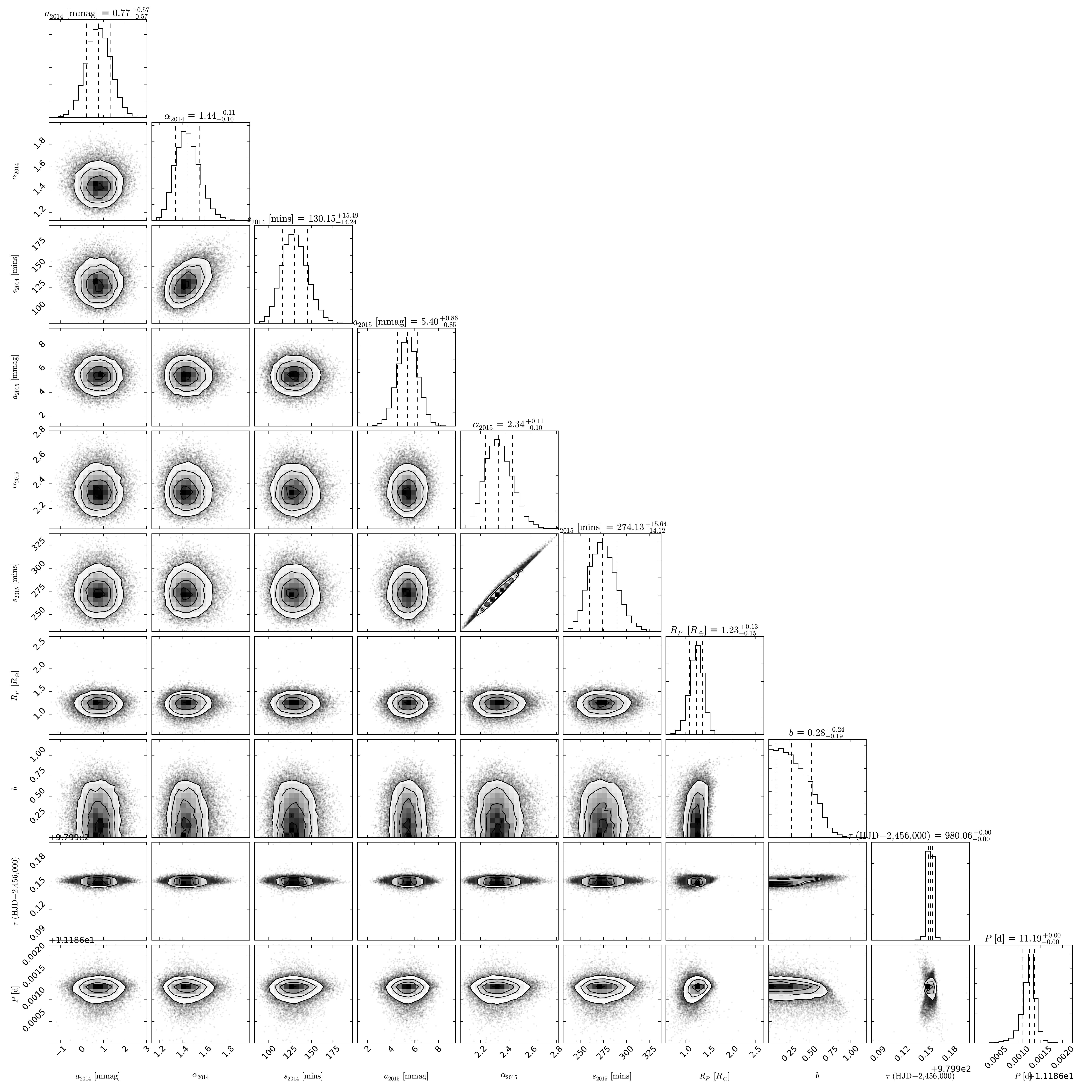}
\caption{
Corner plot of the joint posterior distribution of the fitted parameters
retrieved for model $\mathcal{M}_2$. The parameters are consistent with
those returned by $\mathcal{M}_3$. 
}
\label{fig:postC}
\end{center}
\end{figure*}

We also note that the impact parameter of the signal is non-grazing (see 
Figure~\ref{fig:postC}). This is important because observational bias of the
transit method, given the \forecaster\ size prediction, makes it less likely
a detected signal would be caught on the limb. Integrating the conditional
probability distribution of \citet{sandford}, we are able to estimate that
it is in fact 35 times more likely a real signal would be non-grazing that
grazing.

In conclusion, analysis of signal C shows it be compatible with that expected
if Proxima b were observed to transit. Whilst promising, this in itself does
not prove the signal is genuinely the transit of Proxima b, however, for
reasons discussed earlier in Section~\ref{sub:badlike}.

\section{TYPE I \& II ERROR RATES}
\label{sec:tests}

\subsection{Evaluating the False-Positive (Type I) Rate}
\label{sub:FPs}

In Section~\ref{sub:badlike}, it was established that the Bayesian evidence
may not be a fully reliable tool for model selection in our case. Accordingly,
we seek alternative methods to interpret the significance of signal C. Whilst
cross-validation would be a powerful alternative, Proxima b's period means that
that only two transits of signal C occur in our \MOST\ photometry and ignoring 
some fraction of the data is undesirable. Instead, we elected to perform
a bootstrapping procedure to emulate our detection approach in the
presence/absence of an injected signal.

We first consider the case of type I errors, which is the most critical term in
assessing the credibility of any signals inferred by our approach. We 
specifically considered evaluating the type I error of model $\mathcal{M}_2$, 
which uses an informatively priored ephemeris but uninformative priors on the 
radius of Proxima b (see Table~\ref{tab:priors}). To do this, we need a set of
representative, synthetic fits of null data.

Since the GP model is argued to not represent a complete noise model (see
Section~\ref{sub:badlike}), we cannot use the GP to generate synthetic, 
representative data sets. We also cannot randomly scramble the original data
to create fake data, as this would remove the time-correlated noise structure
clearly seen in our data. Performing a search for inverted-transits is
also not useful, since flares are asymmetric flux increases mimiccing
such events. The best option remaining is to move along the
data in a rolling-window style.

Accordingly, we use the original, unmodified \MOST\ time series and simply
modify the priors used. Specifically, in 100 fits, we iteratively translate
the prior on $\tau$ by $0.01P$ until we loop back round to the original phase
in the $100^{\mathrm{th}}$ trial. The disadvantage of this
approach is that we can't ensure the null data is actually absent of signal
(in fact, we already have established the presence of a spurious signal in
the form of signal S; see Section~\ref{sub:signalS}). Consequently, our 
false-positive rate estimate may be an overestimate if latent but genuine
transit signals reside in our \MOST\ photometry.

In each fit, we re-run an identical fit as before, using \multi\ to explore
the transit parameters and GP hyper-parameters using otherwise identical 
priors. As before, we also compute the marginal likelihood. These tests, and
the others needed to evaluate the type II error rate, demanded significant
computational resources of tens of thousands of core hours on the NASA PLEIADES
supercomputer and is why we are practically limited to only running 100 such
tests.

\begin{figure*}
\begin{center}
\includegraphics[width=17.0cm,angle=0,clip=true]{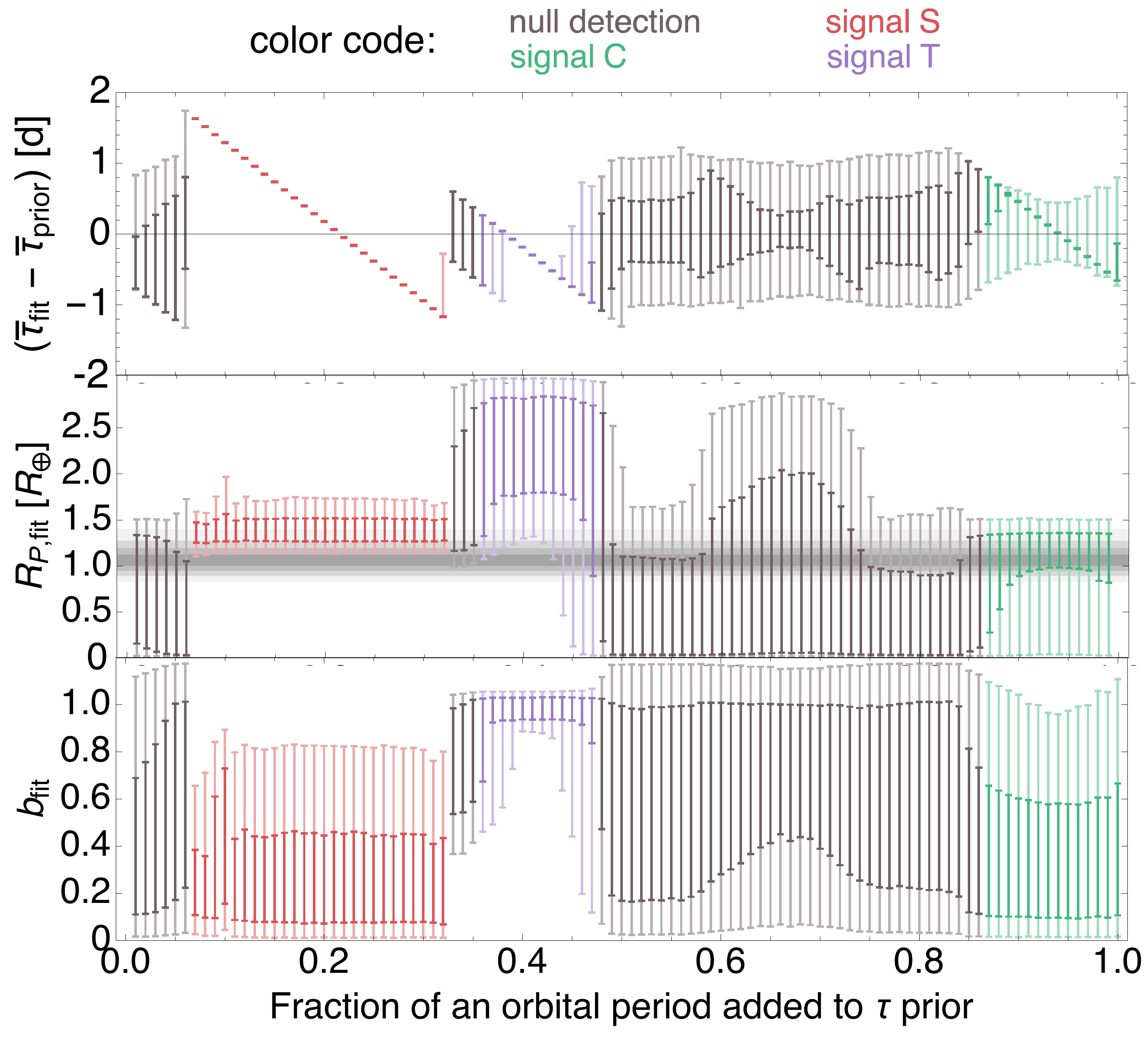}
\caption{
1 (dark bars) and 2 (light bars) $\sigma$ credible intervals from 100 model
$\mathcal{M}_2$ fits on the original \MOST\ data but slightly perturbing the
informative prior on $\tau$ ($x$-axis). Color code key is at the top.
The horizontal region on the second panel denotes the \forecaster\ prediction
for the planetary radius \citep{chen}.
}
\label{fig:FP_params}
\end{center}
\end{figure*}

In order to calculate the false-positive rate, we need to define what
constitutes a ``detection''. A useful metric is to inspect the convergence of
the time of transit minimum posterior. Detections will have a narrow posterior
with most of the density located on a single mode. In contrast, the $\tau$
posterior of null detections will broadly reproduce the prior or display a
broad, unconverged and structured form.

To more explicitly define what we mean by a ``narrow'' posterior, we demand
that the range of the central 50\% quantile, thereby constituting the
majority of the samples is less than one half of the characteristic transit
duration, $\mathbb{T}$. Assuming a circular orbit, adopting a representative
impact parameter\footnote{We use the median of the probability distribution
of $b$ after accounting for observational bias derived in \citet{sandford}.}
of $b=0.5$ and using the mode of our stellar density prior (see 
Table~\ref{tab:priors}) gives a characteristic transit duration of 
$\mathbb{T}=65.3$\,minutes. Further, we require that the Bayes factor between
the fit and the null model $(\mathcal{M}_0)$ exceeds $e$. We therefore require
that:

\begin{itemize}
\item[{\textbf{[i]}}] The interval from the 25\% quantile to 75\% quantile
of the $\tau$ posterior to be less than one half of the characteristic transit
duration, $\mathbb{T}/2$.
\item[{\textbf{[ii]}}] The evidence ratio satisfies $\Delta\log\mathcal{Z}>1$.
\end{itemize}

Turning to the results, we first note that from inspecting the
$\tau$ posteriors of the 100 fits ran, it was immediately obvious that a 
considerable fraction of the fits recovered the signal S and C discussed in
Section~\ref{sub:signalS} \& \ref{sub:signalC}. This can be seen from
Figure~\ref{fig:FP_params}, upper panel, where the $\tau$ posteriors latch onto
a single solution and exhibit a linear-like trend on signals S and C. Since the
$\tau$ prior (x-axis) is shifted each time but the best-fitting solution is the
same, this creates the linear structure observed.

As a result of this behavior, it is necessary to establish a criterion
to identify fits which recovered previously recognized signals, namely signals 
S and C. We define such fits as those for which the median posterior $\tau$ 
sample lies within $\pm0.5$ transit durations of signal S/C's median posterior
$\tau$ sample. Note that here it is unnecessary to use the characteristic 
duration, but instead we can use the actual duration measured from our earlier
model fits. We thus define spurious signals as those satisfying this and also 
\textbf{[i]} \& \textbf{[ii]} (to remove unconverged cases):

\begin{itemize}
\item[{\textbf{[iii]}}] The median posterior $\tau$ sample of the fit is within
$\pm0.5$ transit durations of the median posterior $\tau$ sample of either
signal S or C.
\end{itemize}

\begin{figure*}
\begin{center}
\includegraphics[height=5.75cm,angle=0,clip=true]{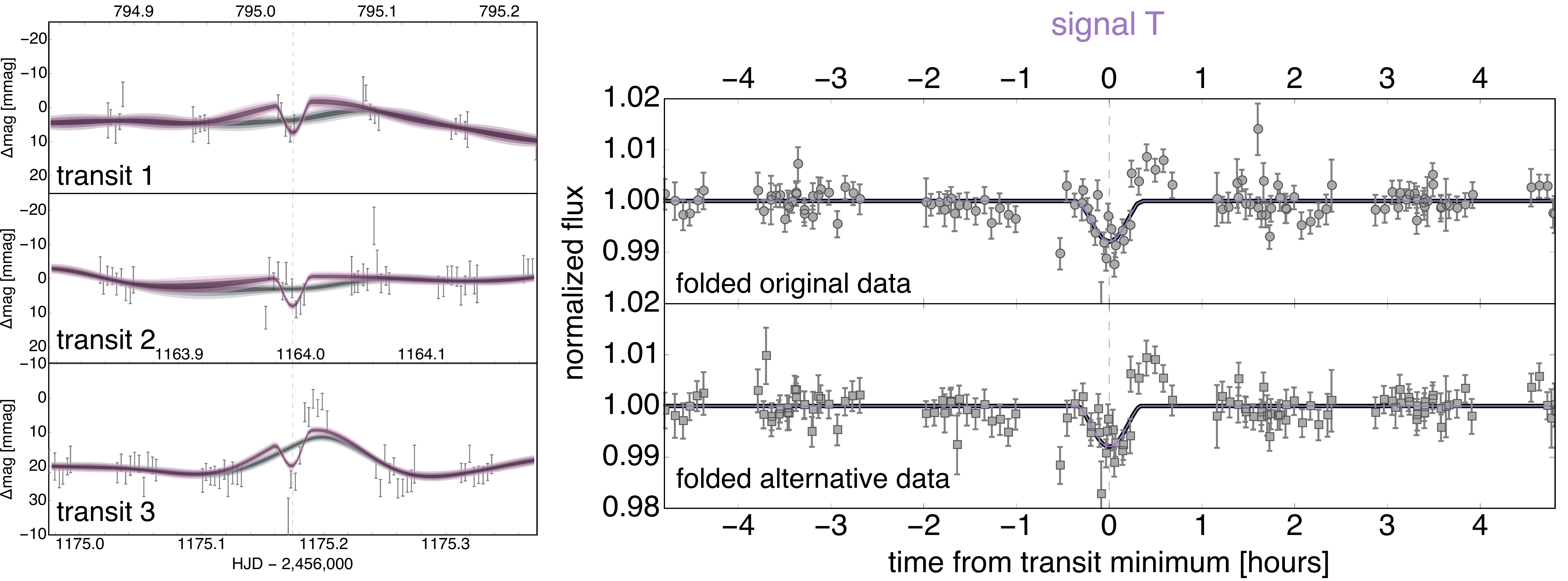}
\caption{
Same as Figure~\ref{fig:signalS} except for the third transit signal found,
signal T. The signal is far less convincing than than signals S and C.
}
\label{fig:signalT}
\end{center}
\end{figure*}

Using this criterion, we found that 26 of the 100 fits converged to signal S
and another 14 converged to signal C. Inspection of Figure~\ref{fig:FP_params}
reveals that a third signal appears to exist in the data, located around a
phase of 0.4, causing 12 additional tests to converge to the same solution.
We label this new signal as signal T. We show the credible
intervals of the fitted parameters $\tau$, $p$ (converted to planetary radii)
and $b$ in Figure~\ref{fig:FP_params}, with each realization color coded to
the aforementioned identities. We also plot the maximum a-posteriori solution 
for signal T in Figure~\ref{fig:signalT}.

Signal T immediately raises skepticism about its validity. Over 95\% of the
posterior trials correspond to a grazing geometry, as evident from the shape
of signal T in Figure~\ref{fig:signalT}. As discussed earlier in 
Section~\ref{sub:signalC}, observational biases mean that it is 35 times more
likely a detected transit would be non-grazing rather grazing. Moreover, it
is generally easier for sharp data artifacts to mimic a V-shaped event than a 
full transit morphology, by virtue of the former's simpler shape. The
ratio-of-radii posterior is pushed up against the upper bounds of the prior,
favoring a planet of $2.35_{-0.55}^{+0.48}$\,$R_{\oplus}$, which is highly
incompatible with the \forecaster\ prediction. Finally, inspection of the
data itself reveals a far less convincing signal than signals C and T. We
assert that signal T would never be genuinely considered a candidate transit
signal of Proxima b, even if the phase had matched with the radial velocity
prediction.

Because signal T would not be considered a detection if it's phase had been
compatible, we not consider is to be a false-positive signal in what follows
and simply discount it from the false-positive evaluations. In contrast, signal
S shows no features that would have caused us to dismiss it as a false-positive,
had it landed at the correct phase. Therefore, of the 74 realizations not 
affected by signals C and T, 26 converge to a single spurious solution. Counting
these cases as unique false-positives would imply a false-positive rate of 
$\mathrm{FPR}=(35\pm7)$\%, whereas counting them as belonging to a single 
false-positive would give $\mathrm{FPR}=(2\pm2)$\%. In conclusion, it is unclear
precisely how to define the false-positive rate from these tests, but certainly
the false-positive rate is non-zero and at least a few percent.

\subsection{Evaluating the False-Negative (Type II) Rate}
\label{sub:FNs}

To evaluate the false-negative rate, we used the same setup as for the type
I tests except we inject a 1.06\,$R_{\oplus}$ planet (see Section
\ref{sub:radius}) with $b=0.5$ into the time series at each phase point.
In each of the 100 tests we ran, the model is seeking an injected Proxima b
like transit signal located within the specified phase prior. Since the
data has been perturbed, it was necessary to re-run the null model, 
$\mathcal{M}_0$, on each of these synthetic data sets, in addition to model
$\mathcal{M}_2$.

We classify null detections (i.e. the false-negatives) as being any case for
which criteria \textbf{i} \& \textbf{ii} are not both satisfied, which
occurs for 23 of our 100 simulations. This sets a minimum limit on the 
false-negative rate of $\mathrm{FNR}=(23\pm5)$\%.

We classify successful recoveries as cases where criteria \textbf{i} \&
\textbf{ii} are both satisfied and that the median of the fit's $\tau$
posterior is less than one half of the injected transit duration from
the the injected transit time. However, we exclude cases where signal
C, S or T is recovered using criterion \textbf{[iii]} (and extended now
to include signal T). Defined in this way, we count 40 successful recoveries.

\begin{figure*}
\begin{center}
\includegraphics[width=17.0cm,angle=0,clip=true]{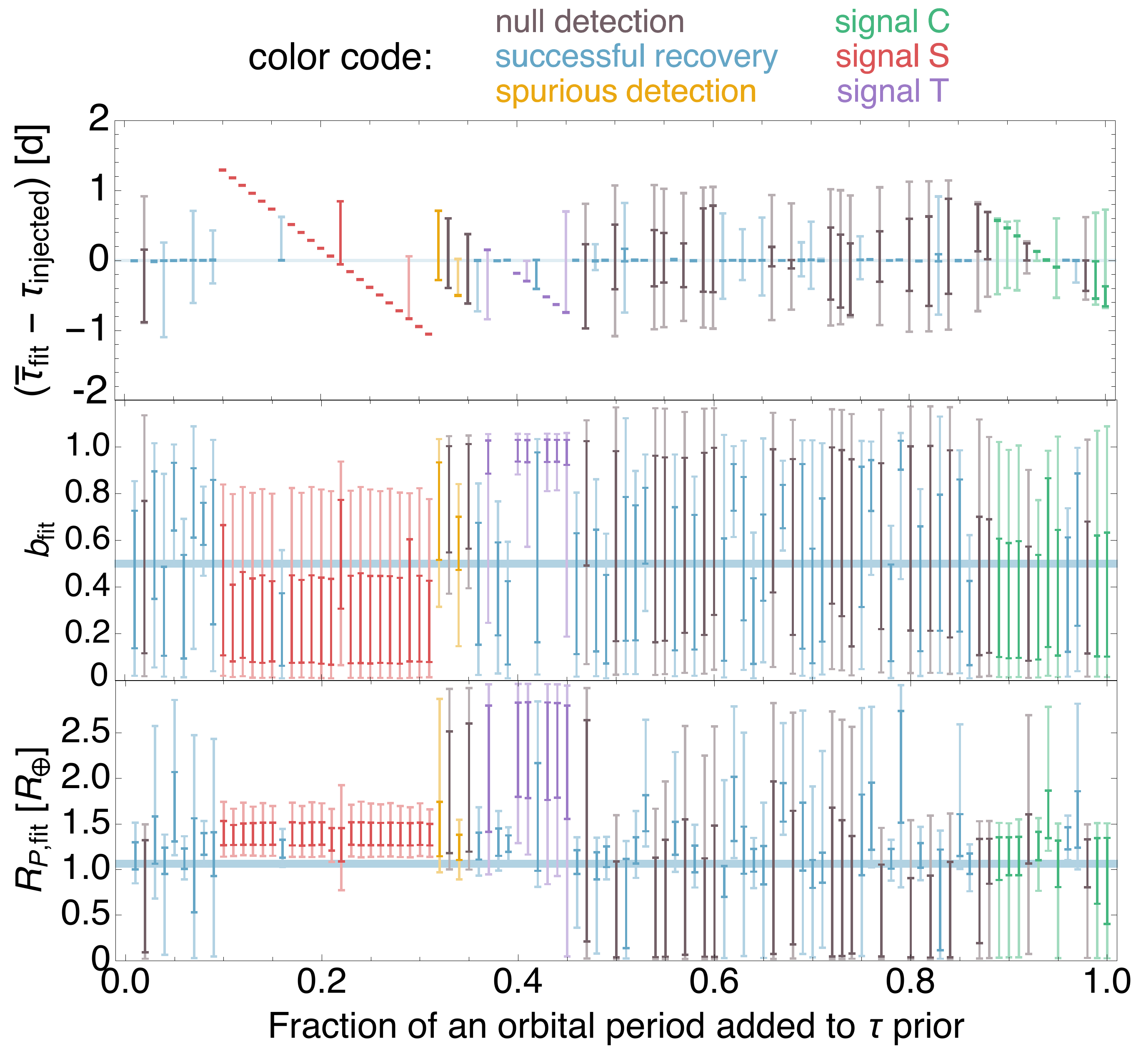}
\caption{
1 (dark bars) and 2 (light bars) $\sigma$ credible intervals from 100 model
$\mathcal{M}_2$ fits on the original \MOST\ data with a Proxima-b-like
transit injected at different phases ($x$-axis). Color coded key is at the top.
The horizontal bars on each plot denote the value of the injected signal.
}
\label{fig:FN_params}
\end{center}
\end{figure*}

In addition to 21 re-detections of signal S, 6 re-detections of signal T
and 8 re-detections of signal C, we find 2 additional ``detections'' which
do not correspond to the injected signal i.e. false-positives. We plot
the credible intervals of these simulations for three key transit parameters
in Figure~\ref{fig:FN_params}, where we color code each of these cases.

Ignoring the previous re-detections, leaves 65 simulations, of which 2 are
false-positives and 40 are successful recoveries. First, this supports the 
previous argument for a false-positive rate of a few percent 
(here $(3\pm3)$\%). Second, it implies that the false-negative rate may be as
high as $\mathrm{FNR}=(39\pm8)$\%.

\subsection{Summary}

Since we lack a complete model to describe the noise structure of our data,
new synthetic, representative data sets cannot be generated to evaluate the 
false-positive and negative rate of our putative Proxima b transit (signal C).
This limits our options to using the original data itself, which is
unfortunately contaminated by three signals - the putative signal itself
and two additional, likely spurious, signals.

Although these signals severely impede our ability to investigate the error
rates of signal C, we estimate that the false-positive rate is at least a few
percent, whereas the false-negative rate is considerably higher at 
$\sim20$-$40\%$. Both of these numbers are sufficiently high to warrant serious
skepticism regarding the reality of signal C. 

At this point, we concluded that the \MOST\ data alone were unable to 
conclusively confirm or reject this candidate signal. Whilst \MOST\ data has
some unique challenges due to the orbital motion and data sparsity, we consider
that the most likely reason why this analysis is so challenging is not 
associated with \MOST\ itself but rather with Proxima's high flare activity, 
which leads to likelihood mis-specification.

Cross-validation is perhaps the model selection tool least likely to suffer from
the affects of likelihood mis-specification. As mentioned earlier, this is
impractical with just two transits observed by \MOST\ for signal C. However,
cross-validating the signal against other data sets would be a viable and
robust way to establish the reality of signal C. Accordingly, we discuss such a
test in the next section.

\section{CROSS-VALIDATING WITH HATSouth DATA}
\label{sec:hats}

\subsection{Observations}
\label{sub:hatsobs}

Independent of the MOST observations, Proxima Cen was also monitored by the 
\HATS\ ground-based telescope network \citep{bakos:2013:hatsouth}. The network
consists of six wide-field photometric instruments located at three 
observatories in the Southern hemisphere (Las Campanas Observatory (LCO) in 
Chile, the High Energy Stereoscopic System (HESS) site in Namibia, and Siding 
Spring Observatory (SSO) in Australia) with two instruments per site. Each 
instrument consists of four 18\,cm diameter astrographs and associated 
4K$\times$4K backside-illuminated CCD cameras and Sloan $r$-band filters, 
placed on a common robotic mount. The four astrographs and cameras together 
cover a $8.2^{\circ} \times 8.2^{\circ}$ mosaic field of view at a pixel scale 
of $3\farcs7$\,pixel$^{-1}$.

Observations of a field containing Proxima Cen were collected as part of the 
general HATSouth transit survey, with a total of 11071\footnote{This number 
does not count observations that were rejected as not useful for high precision 
photometry, or which produced large amplitude outliers in the Proxima Cen light
curve.} composite $3 \times 80$\,s exposures gathered between 2012 June 14 and 
2014 September 20.  These include 3430 observations made with the HS-2 unit at 
LCO, 4630 observations made with the HS-4 unit at the HESS site, and 3011 
observations made with the HS-6 unit at the SSO site. Due to weather, and other
factors, the cadence was non uniform. The median time difference between 
consecutive observations in the full time series is 368\,s.

The data were reduced to trend-filtered light curves using the aperture 
photometry pipeline described by \citet{penev:2013:hats1} and making use of the
External Parameter Decorrelation (EPD) procedure described by 
\citet{bakos:2010:hat11} and the Trend Filtering Algorithm (TFA) due to 
\citet{kovacs:2005:TFA}. One notable change with respect to the procedure 
described by \citet{penev:2013:hats1} is that we made use of the 
proper-motion-corrected positions of celestial sources from the UCAC4 catalog 
\citep{zacharias:2013} to determine the astrometric solution for each image and 
to position the photometric apertures. This modification was essential for 
Proxima Cen which, at the start of the HATSouth observations, was displaced by 
48\arcsec\ (13 pixels) from its J2000.0 location, and moved a total of 
$8\farcs7$ (2.4 pixels) over the 828 days spanned by the observations.  
In this paper we use these data solely to cross-validate the candidate transit 
signals seen in the MOST light curve. The stellar variability of Proxima Cen 
as revealed by HATSouth, and a general search for transits in its light curve,
will be discussed elsewhere. The EPS and TFA processed data are made available
in Table~\ref{tab:hatstable}.

\begin{table}
\caption{Reduced \HATS\ photometry used in this work, after correction for systematic trends by
TFA.
A portion of the table is shown here, the full version is available in the electronic version
of the paper and at \href{\urllink}{this URL}.
} 
\centering 
\begin{tabular}{c c c} 
\hline\hline
HJD$_{\mathrm{UTC}}$ - 2,451,545 & TFA Magnitude & Uncertainty \\ [0.5ex] 
\hline
4547.6443629	& 7.17666 & 0.0018 \\
4547.6501446	& 7.18168 & 0.0018 \\
4595.5309943	& 7.22295 & 0.0016 \\
4598.5210661	& 7.20900 & 0.0017 \\
4598.5250688	& 7.22170 & 0.0014 \\
4598.5290578	& 7.23077 & 0.0014 \\
4598.5333811	& 7.22926 & 0.0015 \\
4598.5360740	& 7.23682 & 0.0027 \\
\vdots & \vdots & \vdots \\
\hline 
\end{tabular}
\label{tab:hatstable} 
\end{table}

\begin{figure*}
\begin{center}
\includegraphics[width=17.0cm,angle=0,clip=true]{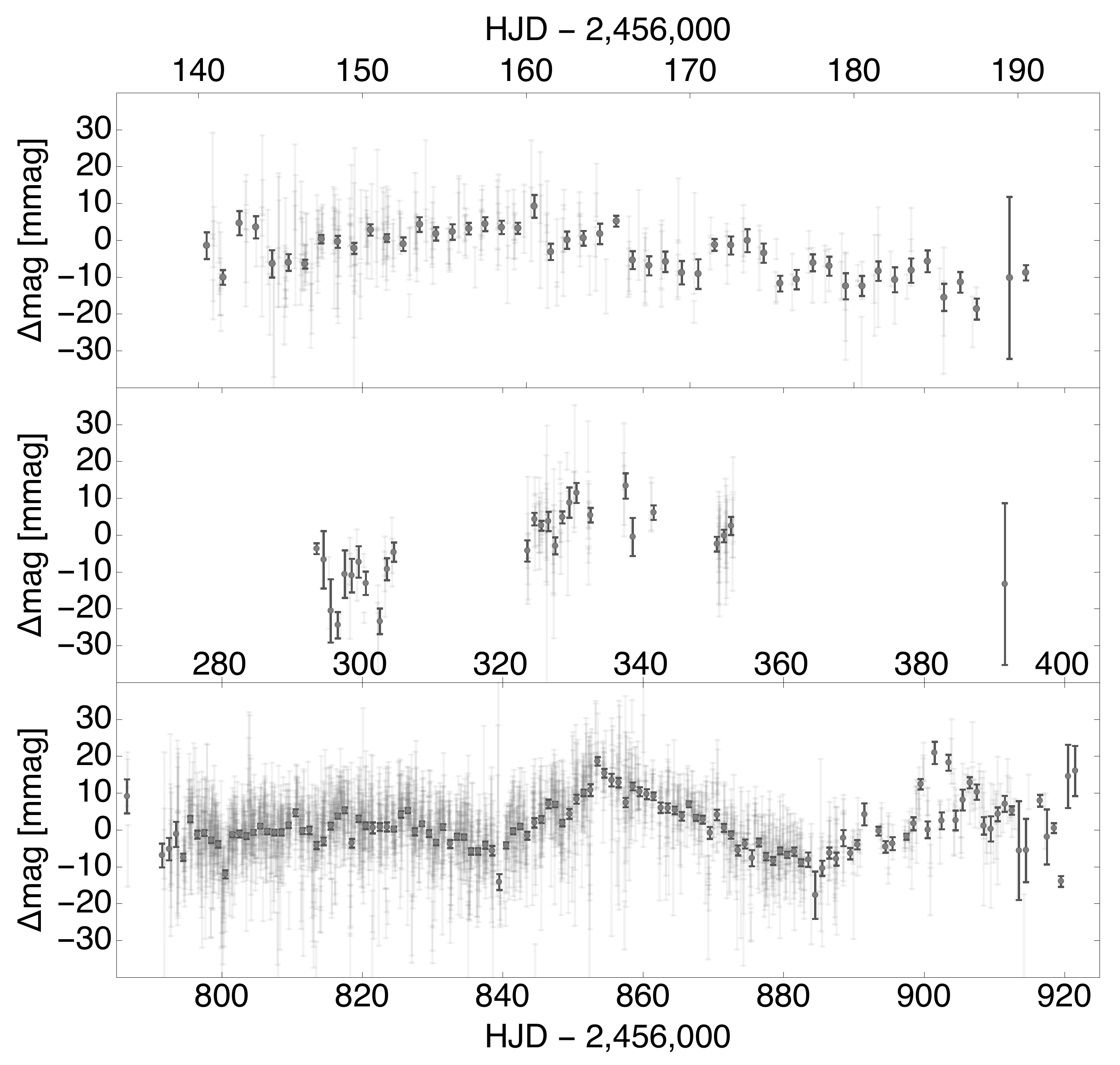}
\caption{
TFA-corrected photometric observations of Proxima with \HATS\ from three
observing seasons shown at native 240\,s cadence (gray bars) and nightly
averages (circles). The TFA procedure suppresses real low-frequency 
variability in addition to removing instrumental trends. The variation seen
at $HJD - 2456850$ is the residual, after TFA, of a much larger 79\,mmag
low-frequency astrophysical variation caused by the rotation of star spots.
}
\label{fig:hats_data}
\end{center}
\end{figure*}

We passed the TFA light curve through a median filter to remove 4\,$\sigma$ 
outliers and the final photometric light curve contains 10869 data points
spanning 206 nights (see Figure~\ref{fig:hats_data}). We find that the
formal measurement uncertainties greatly underestimate the observed scatter
with $\sqrt{\chi^2/n}=7.3$ and thus, in what follows, we treat these 
uncertainties as relative weights but not as reliable estimates of the 
uncertainty on each point.

We note that the light curve of Proxima Cen shows higher scatter at short time
scales than most stars observed by \HATS\ in the same field with comparable 
brightness. Stars within $0.1$\,mag of Proxima Cen have a median post-TFA 
r.m.s. at the same cadence of 5.3\,mmag. For comparison Proxima Cen has an
r.m.s. of 13.4\,mmag without the additional 4\,sigma clipping and median 
filtering, or 10.1\,mmag after these additional cleaning procedures are 
applied. This high scatter is likely astrophysical in nature, and may be 
attributable to rapid low-amplitude variations in brightness due to magnetic 
activity such as flaring.

\subsection{Detrending}

In order to look for the signal C transit, the TFA light curve requires
detrending. We initially attempted to use a Gaussian process, as was used for
the \MOST\ data and experimentation with different kernels again favored the
Mat\'ern-3/2 with a characteristic covariance time scale of 
$s=5.6\pm0.4$\,hours.

Unlike with the analysis of the \MOST\ data though, we are not attempting to
conduct a joint fit of a GP+transit model but rather simply wish to detrend the
light curve using a GP-only model. This difference is important because GPs
are highly flexible and as a detrending tool can actually remove the signals
we seek.

In our case, we wish to fold the detrended data upon signal C's ephemeris and
look for coherent signal. The GP, particularly with a covariance time scale of
just a few hours, is sufficiently flexible to detrend both the long-term changes
and potential transits themselves. Indeed, we can verify this is true since we 
initially attempted to perform the cross-validation using the GP detrended curves
but were unable to recover any injected signals in 100 attempts.

Instead, we used a more tried and tested method for \HATS\ data: median
filtering. Based on previous experience with \HATS\ data, a 2\,day window was
selected. By injecting fake transits into the \HATS\ data and applying the
median filtering, we found that 2\,days did not remove the injected events.
In contrast, when we set the median filtering time scale to 6-12\,hours,
we observed many injected signals were not recovered. For these reasons, we
ultimately settled on the 2\,day window.\\ \\

\subsection{Cross-validation}

\begin{figure*}
\begin{center}
\includegraphics[width=17.0cm,angle=0,clip=true]{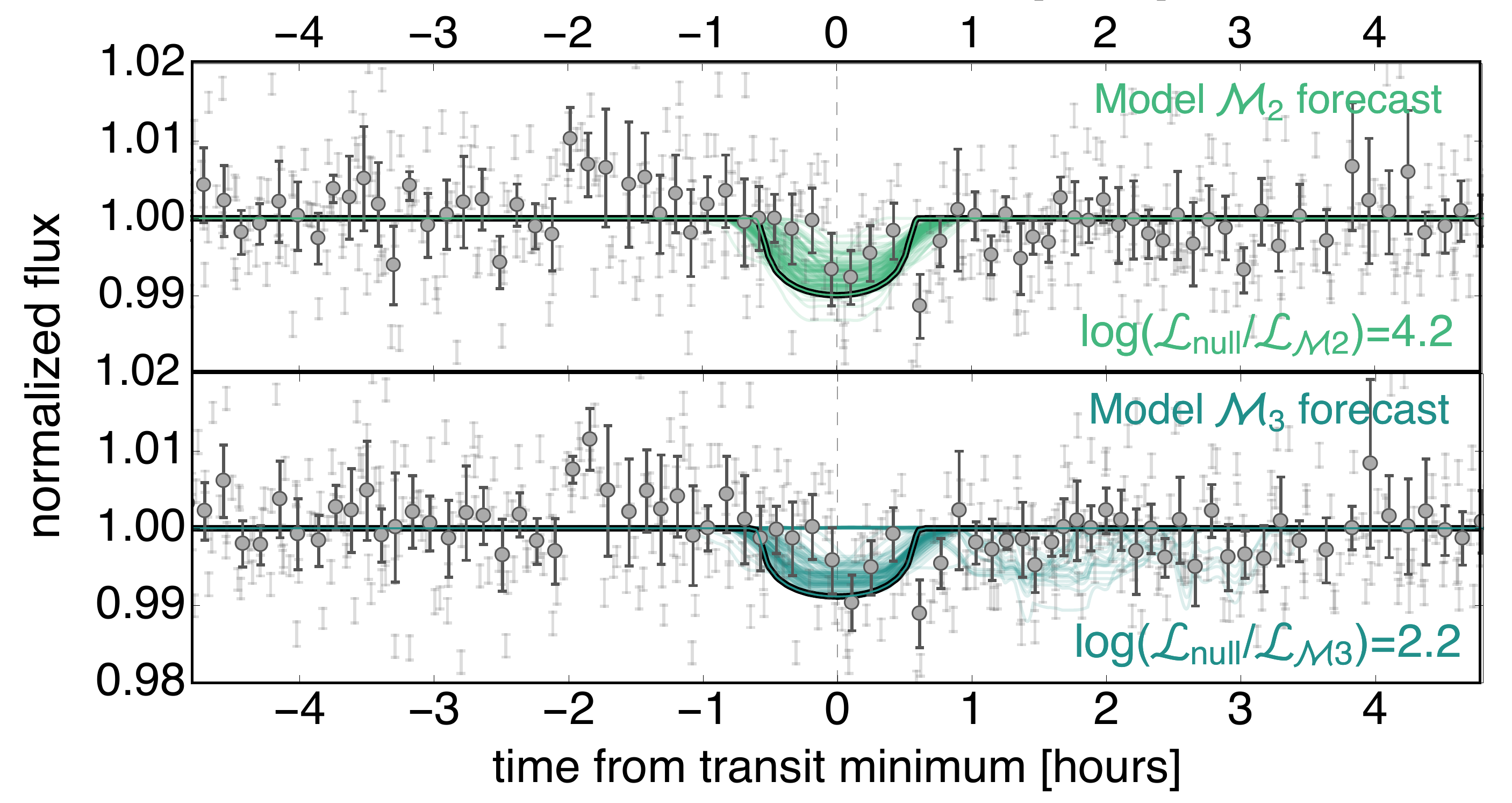}
\caption{
Phase-folded light curve of the detrended \HATS\ data on the maximum
a-posteriori ephemeris from model $\mathcal{M}_2$ (upper) and 
$\mathcal{M}_3$ (lower), corresponding to signal C.
The \HATS\ data does not favor a model with signal C included.
}
\label{fig:hats_fold}
\end{center}
\end{figure*}

Phase-folding the detrended \HATS\ light curve on the maximum a-posteriori
ephemeris of signal C reveals visually evident disagreement between the
expected model and the observations, as shown in Figure~\ref{fig:hats_fold}.
Note that the phase-folded light curve modifies the limb darkening coefficients
to account for the fact \HATS\ is nearly a Sloan r' bandpass (we used
the nearest $T_{\mathrm{eff}}$-$\log g$ grid entry from the tabulated list
of limb darkening coefficients in \citet{2004A+A...428.1001C} for the PHOENIX
stellar atmosphere results).

We binned the phase-folded light curve of 10869 points to 6-point bins in
phase and computed the weighted mean and weighted standard deviation on
each bin. This process enables us to compute realistic uncertainties on each
binned point, to overcome the fact the formal uncertainties on the unbinned
data are underestimates. Against a simple flat line model though, the 1806 
binned points have a $\chi^2=2966$, indicating these weighted standard 
deviation still somewhat underestimate the true uncertainty. Accordingly, we
scale them up by a factor of 1.645 such that $\chi^2=n$.

The $\chi^2$ of the maximum a-posteriori solution of
model $\mathcal{M}_2$ should be lower than a simple flat line, if the transit
were real, but instead it is slightly higher at 
$\chi_{\mathcal{M}2}^2-\chi_0^2=8.4$ for 1803 binned points. As a likelihood 
ratio, this corresponds to a 2.4\,$\sigma$ preference for the null model.
Repeating for $\mathcal{M}_3$ reveals a similar situation, with 
$\chi_{\mathcal{M}3}^2-\chi_0^2=4.4$, or a 1.6\,$\sigma$ preference for the 
null model. The slight dip seen around the time of transit minimum i
Figure~\ref{fig:hats_fold} can be understood as due to autocorrelation, which
detect a very significant $p$-value for in the phase-folded data.

\begin{figure}
\begin{center}
\includegraphics[width=8.4cm,angle=0,clip=true]{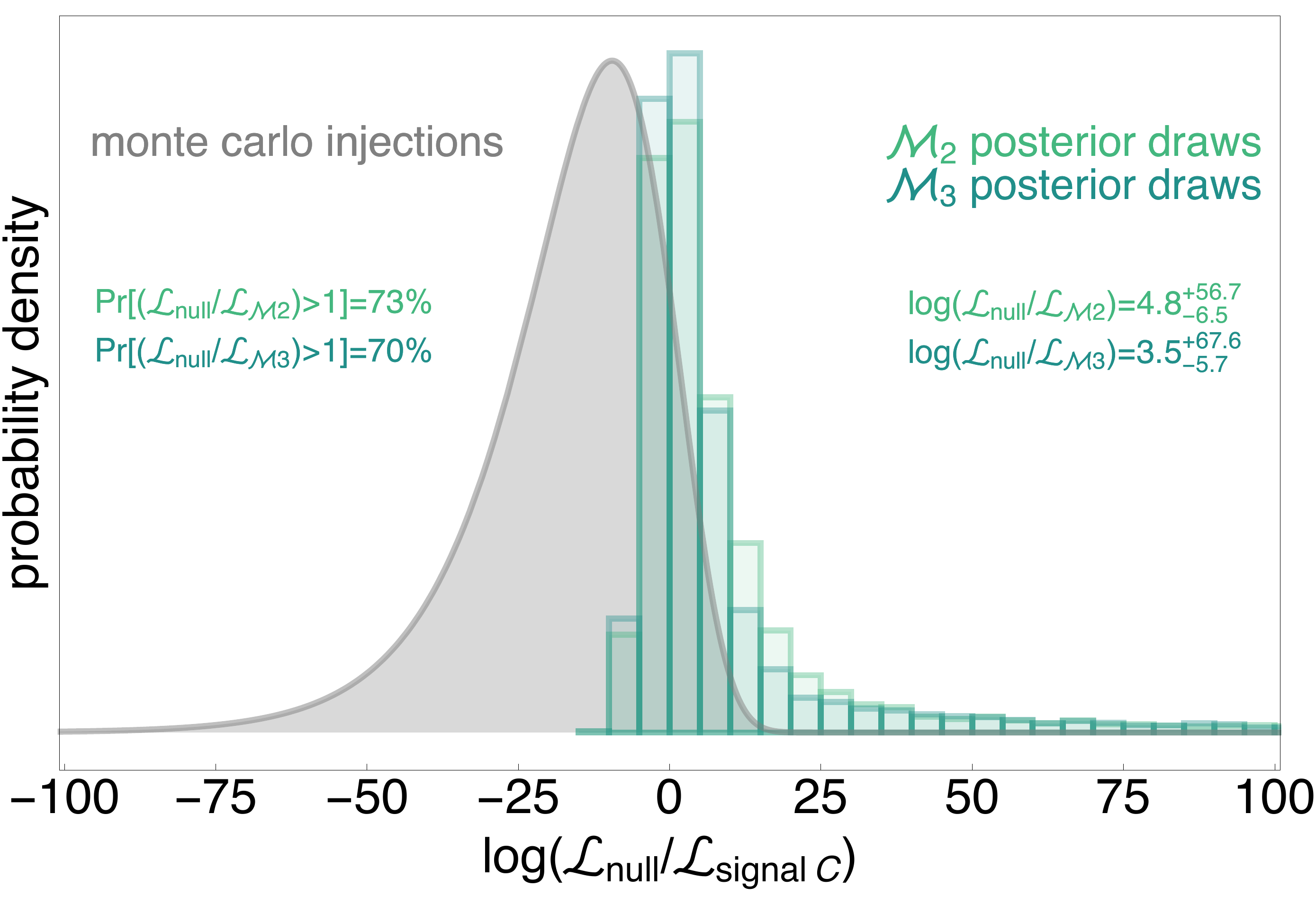}
\caption{
Histogram of the likelihood ratio between a null model and the phase-folded
transit of signal C conditioned on the \HATS\ data using the posterior samples
from models $\mathcal{M}_2$ (light green) and $\mathcal{M}_3$ (dark green).
The gray smoothed histogram shows the distribution from 100 injections of
a signal-C-like planet phase-folded upon the injected ephemeris.
}
\label{fig:hats_histo}
\end{center}
\end{figure}

However, this single realization does not capture the range of plausible models
allowed by models $\mathcal{M}_2$ \& $\mathcal{M}_3$. To assess this, we took
each posterior sample, performed a phase fold and re-binning of the unbinned
data, and then calculated the $\chi^2$ of each model draw relative to the
phase-binned \HATS\ data. At each trial, we re-normalized the errors such that
a flat line through the data yielded a $\chi^2$ equal to the number of data
points, in order to fairly reproduce the procedure described earlier.
The resulting distributions in $\Delta\log\mathcal{L}=-\Delta\chi^2/2$ for
models $\mathcal{M}_2$ and $\mathcal{M}_3$ are shown in 
Figure~\ref{fig:hats_histo}.

Figure~\ref{fig:hats_histo} reveals that signal C is clearly not confirmed by
this cross-validation exercise. Indeed, for both $\mathcal{M}_2$ and
$\mathcal{M}_3$, the majority of the draws favor a null model over the transit
model (73\% of samples for $\mathcal{M}_2$; 70\% of samples for 
$\mathcal{M}_3$).

To provide some context, we injected 100 signal C-like transits
into the \HATS\ at equally spaced phases and found greater
$\Delta\log\mathcal{L}$ values than those observed even in the tail of signal
C's predictions.

We conclude that the \HATS\ data do not provide compelling supporting evidence
for the existence of signal C and moderately disfavor it's existence at the
1-2\,$\sigma$ level. Excluding the signal to greater confidence will likely
require far-red, near-infrared, or infrared photometry, such as that which 
could be provided by \textit{MEarth} or \textit{Spitzer}. Given the sizable 
false-positive/negative rate of the \MOST\ data, the high rate and amplitude of
flares produced by Proxima, the a-priori low transit probability (1.5\%) and 
the lack of support from independent ground-based photometry, we conclude that 
there is not a compelling case to be made that our best candidate signal for a 
transit of Proxima b is genuine.

\section{DISCUSSION \& CONCLUSIONS}
\label{sec:discussion}

In this work, we have searched for the transits of the recently announced 
planetary candidate, Proxima b \citep{nature}, in two seasons (2014 \& 2015) of
space-based photometry obtained using the \MOST\ satellite. Proxima b has an 
a-priori transit probability of 1.5\%, an expected depth of 5\,mmag lasting up 
to an hour \citep{nature}. Accordingly, the signal, should it exist, was 
expected to be quite detectable with our photometry, which after detrending 
exhibits an RMS of 2-3\,mmag per minute with over 15,000 photometric points 
spanning 43.5\,days at a duty cycle of $\sim30$\%.

Proxima Centauri exhibits a few percent level photometric variability in the
\MOST\ bandpass and displays dozens of detectable flares \citep{davenport}.
After removing obvious flares detected with the \FBEYE\ approach
\citep{2014ApJ...797..122D}, we still expect a much greater number of smaller
flares to exist in our data. Indeed, \citet{davenport} predict 5\,mmag flares
(the expected transit depth of Proxima b) to occur every 20\,minutes, on 
average. The sheer volume of such large flare events greatly complicates our
analysis and we argue in this work that even our preferred model for the
time correlated variability, a Gaussian process (GP) with a Mat\'ern-3/2 kernel,
is unlikely to be an accurate description of the star's true behavior.

We conduct Bayesian model selection of a GP-only versus GP+transit model using
the multimodal nested sampling algorithm \multi\ \citep{2008MNRAS.384..449F,
2009MNRAS.398.1601F}. However, the mis-specified likelihood function formally
invalidates the Bayes factors which result. Nevertheless, if we assume that the
sign of the Bayes factor is correct and impose an informative prior on the
period and transit mid-time based of the radial velocity fits of \citet{nature},
then we do find a candidate transit signal for Proxima b in our data, which we
label as signal C.

Signal C's freely fitted planetary radius is consistent with that expected
based on the empirical, probabilistic mass-radius relation of \citet{chen} using
the \forecaster\ package, at $1.23_{-0.15}^{+0.13}$\,$R_{\oplus}$. As expected,
repeating the fits but imposing an extra informative prior on the radius using
\forecaster\ recovers the same signal. However, when we relax the transit
mid-time (or phase) informative prior, a stronger transit signal is detected
at a phase incompatible (at 4\,$\sigma$) with that expected from the radial
velocities, which we label as signal S.

Since our noise model is argued to be inaccurate, we are unable to generate
synthetic data sets for false-positive/negative tests. Instead, we use
the original data and perturb the original fit's phase prior 100 times and
repeat the fit. These tests reveal a false-positive rate of at least a few
percent and a much higher false-negative rate of 20-40\%. This process is
complicated by the presence of signal C and S in the data set though, as
well an additional signal detected during these tests dubbed signal T,
which is likely spurious based on it's V-shaped morphology.

To resolve the validity of signal C, we phase-folded \HATS\ photometry onto
the best-fitted ephemeris of our model and a flat-line provides a slightly
preferred description at 1-2\,$\sigma$ significance. Repeating for the
posterior draws of our model fit reveals that $\sim75$\% of our model
predictions give a worse likelihood than a simple flat-line model through
the \HATS\ data. This final test leads to conclude our signal C is unlikely
to be associated with Proxima b and is most likely a spurious detection
driven by the time correlated noise structure of our data. As a result of
the high false-positive and -negative rates, even when the period of the
signal is known, we did not pursue a blind-period search on this data set,
since the reliability of any ``detections'' would be highly doubtful.

The high flare activity of Proxima Centauri poses a serious challenge for
any photometric follow-up of Proxima b. Even if Proxima b is detected to
transit, we predict that precise transmission spectroscopy of its atmosphere
would be impacted by the flares. The most promising avenue to photometrically
follow-up Proxima will likely be in the red end of the visible spectrum, or 
with infrared measurements, where the star will not only be brighter ($K=4.4$ 
versus $V=11.1$) but the influence of hot flares should be attenuated. Indeed,
far-red/near-infrared/infrared follow-up of the candidate transit signal 
reported here is recommended to conclusively exclude its existence.

\acknowledgments

Based on data from the MOST satellite, a Canadian Space Agency mission, jointly
operated by Microsatellite Systems Canada Inc. (MSCI; formerly Dynacon Inc.), 
the University of Toronto Institute for Aerospace Studies and the University 
of British Columbia, with the assistance of the University of Vienna.

Based in part on observations from the HATSouth network, operated by a 
collaboration consisting of Princeton University (PU), the Max Planck Institute
f\"ur Astronomie (MPIA), the Australian National University (ANU), and the 
Pontificia Universidad Cat\'olica de Chile (PUC).  The station at Las Campanas 
Observatory (LCO) of the Carnegie Institute is operated by PU in conjunction 
with PUC, the station at the High Energy Spectroscopic Survey (H.E.S.S.) site 
is operated in conjunction with MPIA, and the station at Siding Spring 
Observatory (SSO) is operated jointly with ANU. Development of the HATSouth 
project was funded by NSF MRI grant NSF/AST-0723074, and operations have been 
supported by NASA grants NNX09AB29G and NNX12AH91H. J.H. acknowledges support 
from NASA grant NNX14AE87G.

Resources supporting this work were provided by the NASA High-End Computing
(HEC) Program through the NASA Advanced Supercomputing (NAS) Division at Ames 
Research Center.

This research has made use of the {\tt corner.py} code by Dan Foreman-Mackey at 
\href{http://github.com/dfm/corner.py}{github.com/dfm/corner.py}.

We thank members of the Cool Worlds Lab for helpful conversations in preparing this manuscript.
DMK \& JC acknowledges support from NASA grant NNX15AF09G (NASA ADAP Program).
DBG, JMM, AFJM, SMR acknowledge support from NSERC (Canada). JRAD is supported by
an NSF Astronomy and Astrophysics Postdoctoral Fellowship under award AST-1501418. AJ acknowleges support from FONDECYT project 1130857, BASAL CATA PFB-06, and by the Ministry of the Economy, Development, and Tourism's Programa Iniciativa Cient\'{i}fica Milenio through grant IC\,120009, awarded to the Millenium Institute of Astrophysics (MAS).

\end{document}